%% file: expo9812.tex
\input sacmace

\input epsf

\francmodefalse
\def\r{{\rm r}}

\def\phib{\phi}
\def\z{\tilde z}

\def\MS{{\overline{\rm MS}}}
\def\z{r}
\def\pseudoe{{\tilde \varepsilon}}

\def\cite{\refs}   
\def\eqref{\eqns}   


\def\ttd3{1~}
\def\tteps{2~}
\def\ttn4{3~}
\def\ttdg{4~}
\def\ttds{5~}
\def\tteqst{6~}
\def\ttcramp{7~}
\def\ttcrampb{8~}
\def\ttlg{9~}
\def\ttnic{10~}
\def\ttsok{11~}
\def\ttepsiold{12~}
\def\ttpelvic{13~}
\def\ttht{14~}
\def\ttmc{15~}
\def\tterg{16~}
\def\ttEXP{17~}

\def\free{{\rm free}}
\def\bc{{\rm bc}}
\nref\rLGZJi{J.C. Le Guillou and J. Zinn-Justin, {\it
Phys. Rev. Lett.} 39 (1977) 95.} 
\nref\rLGZJii{J.C. Le Guillou and J. Zinn-Justin,{\it Phys. Rev.} B21 (1980)
3976.} 
\nref\rbook{J. Zinn-Justin, 1989, {\it Quantum Field
Theory and Critical Phenomena}, in particular chap.~28 of third ed., Clarendon
Press (Oxford 1989, third ed. 1996).} 
\nref\reviewLGZJ{{\it Phase
Transitions} vol. B72, M. L\'evy, J.C. Le Guillou and J.~Zinn-Justin eds.
Proceedings of the Carg\`ese Summer Institute 1980 (Plenum, New York
1982).}
\nref\rBLGZJN{E. Br\'ezin, J.C. Le Guillou, J. Zinn-Justin and B.G. Nickel,
{\it Phys. Lett.} 44A (1973) 227.} 
\nref\rParisi{G. Parisi, {\it Carg\`ese Lectures 1973}, published in
{\it J. Stat. Phys.} 23 (1980) 49.}
\nref\rNMB{B.G. Nickel, D.I. Meiron, G.B. Baker, Univ. of Guelph Report 1977.}
\nref\rLGZJiii{J.C. Le Guillou
and J. Zinn-Justin, {\it J. Physique Lett. (Paris)} 46 (1985) L137; {\it J.
Physique (Paris)} 48 (1987) 19; 50 (1989) 1365.}
\nref\rLGZJed{Many articles on this topic
are reprinted in {\it Large Order Behaviour of Perturbation Theory}, {\it
Current Physics} vol.~7, J.C. Le Guillou and J. Zinn-Justin eds.,
(North-Holland, Amsterdam 1990).}

\nref\rBNGMo{G.A. Baker, B.G. Nickel, M.S. Green
and D.I. Meiron, {\it Phys. Rev. Lett.} 36  (1976) 1351.}

\nref\repsilonv{A.A. Vladimirov,
D.I. Kazakov and O.V. Tarasov, {\it Zh. Eksp. Teor. Fiz.} 77 (1979) 1035 ({\it
Sov. Phys. JETP} 50 (1979) 521)\semi 
K.G. Chetyrkin, A.L. Kataev and F.V. Tkachov, {\it Phys. Lett.} B99 (1981)
147; B101 (1981) 457(E)\semi
K.G. Chetyrkin and F.V. Tkachov, {\it Nucl. Phys.} B192 (1981) 159\semi
K.G. Chetyrkin, S.G. Gorishny, S.A. Larin and F.V. Tkachov, {\it Phys. Lett.}
132B (1983) 351\semi
D.I. Kazakov, {\it Phys. Lett.} 133B (1983) 406\semi
S.G. Gorishny, S.A. Larin and F.V. Tkachov, {\it Phys. Lett.} 101A (1984)
120}
\nref\repsigood{H. Kleinert, J. Neu, V. Schulte-Frohlinde, K.G. Chetyrkin and
S.A. Larin, {\it Phys. Lett.} B272 (1991) 39, Erratum B319 (1993) 545.
See also H. Kleinert and V. Schulte-Frohlinde,
Critical Properties of $\Phi^4$-Theories
World-Scientific, 1997.}
\nref\rRGZJ{R.~Guida and J.~Zinn-Justin, Nucl. Phys. B489[FS] (1997) 626.} 
\nref\rLipa{ J.A. Lipa, D.R. Swanson, J. Nissen,
T.C.P. Chui and U.E. Israelson, {\it Phys. Rev. Lett.} 76 (1996) 944.}
\nref\rNickel{B.G. Nickel, {\it Physica} 106A (1981) 48.}
\nref\rZJhts{J. Zinn-Justin, {J. Physique (Paris)} 42 (1981) 783.}
\nref\rAdler{J. Adler, {\it J. Phys.} A16 (1983) 3585.}
\nref\rFiCh{J.H. Chen, M.E. Fisher and B.G. Nickel, {\it Phys. Rev. Lett.}
48 (1982) 630\semi
M.E. Fisher and J.H. Chen, {\it J. Physique (Paris)} 46 (1985) 1645.}
\nref\rGutt{A.J.~Guttmann, {\it J. Phys.} A20 (1987) 1855\semi
A.J.~Guttmann and I.G.~Enting, {\it J. Phys.} A27 (1994) 8007.} 
\nref\rNiRe{B.G. Nickel and J.J. Rehr, {\it J. Stat. Phys.} 61 (1990) 1.}
\nref\rBWW{G.M. Avdeeva and A.A. Migdal, {\it JETP Lett.} 16 (1972) 178\semi
E. Br\'ezin,
D.J. Wallace and K.G. Wilson, {\it Phys. Rev. Lett.} 29 (1972) 591; {\it Phys.
Rev.} B7 (1973) 232.} 
\nref\rWZAN{D.J. Wallace and R.P.K. Zia, {\it J.
Phys. C: Solid State Phys.} 7 (1974) 3480.}
\nref\rNA{J.F. Nicoll and P.C. Albright,
{\it Phys. Rev.} B31 (1985) 4576.}
\nref\rbervil{C. Bervillier, {\it Phys. Rev.} B34 (1986) 8141;}
\nref\rBBMN{C. Bagnuls, C. Bervillier, D.I. Meiron and B.G.
Nickel, {\it Phys. 
Rev.} B35 (1987) 3585.}
\nref\rHaDo{F.J. Halfkann and V. Dohm, {\it Z.
Phys.} B89 (1992) 79.}
\nref\rRAJA{A.K. Rajantie, {\it Nucl. Phys.} B480 (1996) 729,
hep-ph/9606216: In this article the expansion is given in analytic form up
to three loops.} 
\nref\rJoseph{P. Schofield, {\it Phys. Rev. Lett.} 23 (1969) 109\semi 
P. Schofield, J.D.
Litster and J.T. Ho, {\it Phys. Rev. Lett.} 23 (1969) 1098\semi B.D.
Josephson, {\it J. Phys. C: Solid State Phys.} 2 (1969) 1113.}
\nref\rliu{A.J. Liu and M.E. Fisher, {\it Physica} A156 (1989) 35.}
\nref\raharony{A. Aharony and P.C. Hohenberg, {\it Phys. Rev.} B13 (1976)
3081; {\it Physica} 86-88B (1977) 611.}
\nref\rprivman{V. Privman, P.C. Hohenberg, A. Aharony, {\it Universal Critical
Point Amplitude Relations}, in Phase Transitions and Critical Phenomena
vol.~14, C.~Domb and J.L.~Lebowitz eds., (Academic Press 1991).}
\nref\rBorsom{
 J.P. Eckmann, J. Magnen and R. S\'en\'eor, {\it Commun.
Math. Phys.} 39 (1975) 251\semi
J.S. Feldman and K. Osterwalder, {\it Ann. Phys. (NY)} 97 (1976) 80\semi
J. Magnen and R. S\'en\'eor, {\it Commun. Math. Phys.} 56 (1977) 237\semi
J.-P. Eckmann and H. Epstein,  {\it Commun. Math. Phys.} 68 (1979) 245.} 
\nref\rLoef{
Summation of series by Borel transformation and mapping 
was proposed by J.J. Loeffel, {\it Saclay Report}, DPh-T/76/20 unpublished.} 
\nref\rfishertarkoa{M.E. Fisher and H.B. Tarko, {\it Phys. Rev.} B11 (1975)
1131.} 
\nref\rgauntdomb{D.S. Gaunt and C. Domb, {\it J. Phys.C} 3 (1970) 1442.}%
\nref\rAnSo{S.A. Antonenko and A.I. Sokolov, {\it Phys. Rev.} E51 (1995) 1894:
in this article the expressions for the general $O(N)$ theory are reported.
See also A. Sokolov, Fizika Tverdogo Tela 40, N.7 (1998) for related work.}
\nref\rZLFish{
S. Zinn, S.-N. Lai, and M.E. Fisher, {\it Phys. Rev. E} 54 (1996) 1176.}
\nref\rPelVicii{A. Pelissetto and E. Vicari, Nucl. Phys. B519 (1998) 626.}
\nref\rsokolov{
A.I. Sokolov, V.A. Ul'kov and E.V. Orlov, 
{\it Renormalization Group 96}, Third Int. Conf.,Dubna, 1997,
pp.378-383 and A.I. Sokolov, V.A. Ul'kov and E.V. Orlov, J. Phys. Studies,
1 (1997) 362-365.\semi
A.I. Sokolov,{\it Fizika Tverdogo Tela (Solid State
Physics)} 38 (1996) 640.\semi
A.I. Sokolov, V.A. Ul'kov and E.V. Orlov, {\i Phys. Lett. } A227 (1997) 255.
}
\nref\rreisz{T. Reisz,{\it Phys. Lett.} B360 (1995) 77.}
\nref\rbuco{
  P. Butera and M. Comi, {\it Phys. Rev.} E55 (1997) 6391.}
\nref\rtsypin{M.M. Tsypin,{\it Phys. Rev. Lett.} 73 (1994) 2015.}
\nref\rkimlandau{J-K Kim and D.P. Landau, {\it Nucl. Phys. Proc. Suppl.} 
53 (1997) 706, hep-lat/9608072.} 
\nref\rwetterich{N. Tetradis and C. Wetterich, {\it Nucl. Phys.} 
B422 (1994) 541.}
\nref\rwetterichb{J. Berges, N. Tetradis and C. Wetterich,
{\it Phys. Rev. Lett.} 77 (1996) 873.}

\nref\rBGNMU{D.B. Murray and B.G. Nickel, unpublished.
(We have taken central values $g^*=g_0$ (preferred by those authors)
for critical exponents.)}
\nref\rmorris{T. Morris, {\it Nucl. Phys.} B495 (1997) 477, hep-th/9612117.}
\nref\rCot{J.P. Cotton, J. Physique Lett. (Paris) 41 (1980) L231.}
\nref\rBUCOM{P. Butera and M. Comi, {\it Phys. Rev.} B56 (1997) 8212,
hep-lat/9703018.} 
\nref\rMacDo{D. MacDonald, D.L. Hunter, K. Kelly and N. Jan, J. Phys. A25
(1992) 1429.} 
\nref\FeMoW{M. Ferer, M.A. Moore and M. Wortis, Phys. Rev. B8 (1973) 5205.}
\nref\RiFiAl{D.S. Ritchie and M.E. Fisher, Phys. Rev. B5(1972)2668; S.
McKenzie, C. Domb and D.L. Hunter, J. Phys. A15 (1982) 3899; M. Ferer and A.
Hamid-Aidinejad, Phys. Rev. B34 (1986) 6481.}  

\nref\rcacape{S. Caracciolo, M.S. Causo and  A. Pelissetto, 
Phys. Rev. E57 (1998) R1215,   
cond-mat 9703250.}
\nref\rsokal{B. Li, N. Madras and A.D. Sokal, J. Stat. Phys. 
{\bf 80} (1995) 661. }

\nref\rBCGS{G. Bhanot, M.
Creutz,  U. Gl\"assner and K. Schilling, {\it Phys. Rev.} B49 (1994) 12909.} 
\nref\rFeLa{A.M. Ferremberg and D.P. Landau, Phys. Rev. B44 (1991) 5081.}
\nref\rBloet{
H.W.J. Bl\"ote,  J.R. Heringa, A. Hoogland, E.W. Meyer and T.S. Smit,
cond-math 9602020.
\semi
H.W.J. Bl\"ote, A. Compagner, 
J.H. Croockewit, Y.T.J.C. Fonk, J.R. Heringa, A. Hoogland, T.S. Smit and A.L.
van Villingen, {\it Physica} A161 (1989) 1.
\semi
A.L. Talapov and H.W.J. Bl\"ote, {\it J. Phys. A}29(1996)5727.
}
\nref\rRaGu{C.F. Baillie, R. Gupta,
K.A. Hawick and G.S. Pawley, {\it Phys. Rev.} B45 (1992) 10438, and references
therein \semi R. Gupta and P. Tamayo, {\it Critical exponents of the 3D Ising 
model}, LA UR-96-93 preprint, cond-mat/9601048.}
\nref\rKosuzu{M. Kolesik and M. Suzuki, cond-math 9411109.}
\nref\rGHM{A.P. Gottlob, M. Hasenbusch and S. Meyer, {\it Proceedings of the
Int. Symp. on Lattice Field Theory "Lattice93"}, Amsterdam 1992. 
\semi See also M. Hasenbusch and S. Meyer, Phys. Lett. B241,(1990) 238;
A.P. Gottlob and M. Hasenbusch, Physica A 201, (1993) 593. }
\nref\rJanke{W. Janke, Phys.Lett. A148 (1990) 306.}
\nref\rBFMM{H.G. Ballesteros, L.A. Fernandez, V. Martin-Mayor and A.Munoz
Sudupe, Phys. Lett. B387, (1996) 125.
Values for $N=1$ from private communication of L.A.Fernandez.} 
\nref\rJAHO{ C. Holm and W. Janke, Phys. Rev. B48 (1993) 936;
Phys. Lett. A 173 (1993) 8; hep-lat/9605024
\semi
K. Chen, A.M. Ferrenberg and D.P. Landau, Phys. Rev. B48, (1993) 3249;
J. Appl.Phys. 73 (1993) 5488.
}
\nref\rKAKA{K. Kanaya, S. Kaya, Phys. Rev. D51 (1995) 2404.}
\nref\rWRolf{C. Wieczerkowski and J. Rolf, hep-th/9607042, MS-TPI 9610,
NBI-HE-96-34.} 

\newskip\tableskipamount \tableskipamount=8pt plus 3pt minus 3pt
\def\tableskip{\vskip\tableskipamount}
\def\g{\tilde{g}}
\preprint{SPhT-t97/040}

\title{Critical exponents of the $N$-vector model}

\centerline{R.~Guida* and J.~Zinn-Justin**}
\medskip{\it
CEA-Saclay, Service de Physique Th\'eorique***, F-91191 Gif-sur-Yvette
\goodbreak Cedex, FRANCE} 
\footnote{}{${}^{*}$email: guida@spht.saclay.cea.fr}
\footnote{}{${}^{**}$email: zinn@spht.saclay.cea.fr}
\footnote{}{${}^{***}$Laboratoire de la Direction des
Sciences de la Mati\`ere du 
Commissariat \`a l'Energie Atomique}

\abstract
Recently the series for two RG functions (corresponding to the anomalous
dimensions of  the fields $\phi$ and $\phi^2$) of the 3D $\phi^4$ field theory
have been extended to next order (seven loops) by Murray and Nickel. We
examine here the influence of these additional terms on the estimates of
critical exponents of the $N$-vector model, using some new ideas in the
context of the Borel summation techniques.
The estimates have slightly changed, but remain within errors of the previous
evaluation. Exponents like $\eta$ (related to the field anomalous dimension),
which were poorly determined in the previous evaluation of Le
Guillou--Zinn-Justin, have seen their apparent errors significantly decrease.
More importantly, perhaps, summation errors are better determined.\par
The change in  exponents affects the recently determined ratios of amplitudes
and we report the corresponding new values.\par
Finally, because an error has been discovered in the last order of the
published $\varepsilon=4-d$ expansions (order $\varepsilon^5$), we have also
reanalyzed the determination of exponents from the $\varepsilon$-expansion.
\par
The conclusion is that the general agreement between $\varepsilon$-expansion
and 3D series has improved with respect to Le Guillou--Zinn-Justin. 
\endabstract

\vfill\eject
\section Introduction and summary of results

Recently the perturbative expansions of the anomalous dimensions of the fields
$\phi$ and $\phi^2$  for the $O(N)$ symmetric $(\phi^2)^2_{d=3}$ field theory
have been extended to next order (seven loops) in the case $N=0,\cdots 3$
by Murray and Nickel \refs{\rBGNMU}.
This rather impressive result has led us to reexamine the determinations of
the critical exponents for $N=0$ (polymers), $N=1$ (Ising like
systems), $N=2$ (superfluid Helium) and $N=3$ (real ferromagnets). For
completeness we have added results (at six loops) for $N=4$ which correspond
to the Higgs sector of the Standard Model at finite temperature.
A limitation of the present work is that the series for
the RG $\beta$-functions have not been extended (they remain at six loops)
and for several exponents this now is the main source of error.
\par
Critical exponents have also been calculated in the form of $\varepsilon=4-d$
expansions, up to five loops \cite{\repsilonv}. Recently a slight error in the
previously published series has been corrected \cite{\repsigood}, and this has
motivated us to also reexamine the corresponding estimates
(again adding $N=4$ results).
\par
For the reader who is not interested in details we summarize our main results
for $N=0,\cdots 3$ in Table \ttd3 ($d=3$)
and in Table \tteps ($\varepsilon$-expansion) 
while $N=4$ results for both methods can be found in Table \ttn4.
We have chosen central values which satisfy all scaling relations, but the
apparent errors for $\gamma,\nu,\beta,\eta$ in general have been determined
independently. For the $d=3$ IR fixed point value $g^*$ we give results both
in the usual field theory normalization (Eqs.~\egrzerom{}) and in the
normalization used by Nickel \cite{\rNMB}, 
$$\g={N+8\over 48\pi} g$$ 
which is such that the fixed point value is close to $1$.
\par
Note that  in Table \ttd3 in addition to the
plain  $\varepsilon$-expansion results (denoted as "free") 
we report some additional results   
denoted as
"bc" (i.e. with boundary condition) that 
try to incorporate the knowledge of the exact $d=2$ values by
summing the series $(f(\varepsilon)-f(2))/(2-\varepsilon)$, where
$f(\varepsilon)$ is an exponent with known 2D value.
In the case of the exponent $\nu$ for $N=0$ the $d=1$ value is also known.
We have checked that incorporating this additional piece of information has no
significant impact on the final result.\par 
For $N \ge 2$ the analysis of the series with boundary conditions is quite
difficult. Therefore we present here only central values, but no error
estimates. Values and errors of the corresponding free estimates give some
indication.
\par
Let us finally emphasize that we have no real knowledge about the analytic
properties of exponents when $d$ approaches $2$. Therefore the bc
values could be affected by systematic effects.\par
The article then is organized as follows: 
in section \ssefact~we summarize a few ideas about  
perturbative expansion at fixed $d=3$ dimension and $\varepsilon$-expansion. 
In section \ssUQsum~we briefly  recall the Borel summation method based on
a conformal mapping of the complex cut plane. Several new variations of the
practical implementation of the general method are explained.
In section \sssexp~we recall the idea of the pseudo-epsilon expansion and
introduce the exponents' correlation analysis, which consists in eliminating
the coupling constant 
between different exponents. Section \ssnumres~contains a discussion of the
numerical results. Finally the new values of exponents slightly affect the
recently published results \refs{\rRGZJ} for the equation of state of the 3D
Ising model, and we present the new determination in section \ssupdate~
(as well as a revised version of $\varepsilon$-expansion predictions).
%
 \midinsert
$$ \vbox{\elevenpoint\offinterlineskip\tabskip=0pt\halign to \hsize
{& \vrule#\tabskip=0em plus1em & \strut\ # \
& \vrule#& \strut # 
& \vrule#& \strut # 
& \vrule#& \strut # 
& \vrule#& \strut # 
&\vrule#\tabskip=0pt\cr
\noalign{ \centerline{Table \ttd3}\tableskip}
\noalign{\centerline{\it Critical exponents of the
$O(N)$ models from $d=3$ expansion (present work).} \tableskip}
\fileth
height2.0pt& \omit&& \omit&& \omit&&\omit&& \omit&\cr
&$ \hfill N \hfill$&&$ \hfill 0
\hfill$&&$ \hfill 1 \hfill$&&$ \hfill 2
\hfill$&&$ \hfill 3 \hfill$&\cr
height2.0pt& \omit&& \omit&& \omit&& \omit&& \omit&\cr
\fileth
height2.0pt& \omit&& \omit&& \omit&& \omit&& \omit&\cr
&$ \hfill \g^*_{\rm Ni} \hfill$
&&$ \hfill 1.413\pm 0.006  \hfill
$&&$ \hfill 1.411\pm0.004\hfill$
&&$ \hfill 1.403\pm 0.003 \hfill$
&&$\hfill 1.390\pm0.004 \hfill$&\cr
height2.0pt& \omit&& \omit&& \omit&& \omit&& \omit&\cr
&$ \hfill g^* \hfill$
&&$ \hfill  26.63\pm 0.11  \hfill$
&&$ \hfill 23.64\pm0.07 \hfill$
&&$ \hfill 21.16\pm 0.05 \hfill$
&&$\hfill 19.06\pm0.05 \hfill$&\cr 
height2.0pt& \omit&& \omit&& \omit&& \omit&& \omit&\cr
&$ \hfill \gamma \hfill$
&&$ \hfill 1.1596\pm0.0020 \hfill$
&&$ \hfill 1.2396\pm 0.0013\hfill$
&&$ \hfill 1.3169\pm0.0020  \hfill$
&&$\hfill 1.3895\pm0.0050 \hfill$&\cr
height2.0pt& \omit&& \omit&& \omit&& \omit&& \omit&\cr
&$ \hfill \nu \hfill$
&&$ \hfill 0.5882\pm 0.0011 \hfill$
&&$ \hfill 0.6304\pm 0.0013\hfill$
&&$ \hfill 0.6703\pm 0.0015  \hfill$
&&$\hfill 0.7073\pm 0.0035 \hfill$&\cr
height2.0pt& \omit&& \omit&& \omit&& \omit&&\omit&\cr
&$ \hfill \eta \hfill$
&&$ \hfill 0.0284\pm0.0025\hfill$
&&$ \hfill 0.0335\pm0.0025\hfill$
&&$ \hfill 0.0354\pm0.0025  \hfill$
&&$\hfill 0.0355\pm0.0025 \hfill$&\cr
height2.0pt& \omit&& \omit&& \omit&& \omit&& \omit&\cr
&$ \hfill \beta \hfill$
&&$ \hfill 0.3024\pm0.0008\hfill$
&&$ \hfill 0.3258\pm0.0014\hfill$
&&$ \hfill 0.3470\pm0.0016  \hfill$
&&$\hfill 0.3662\pm0.0025 \hfill$&\cr
height2.0pt& \omit&& \omit&& \omit&& \omit&& \omit&\cr
&$ \hfill \alpha \hfill$&&
$ \hfill 0.235\pm0.003 \hfill$
&&$ \hfill 0.109\pm0.004\hfill$
&&$ \hfill -0.011\pm0.004  \hfill$
&&$\hfill -0.122\pm0.010 \hfill$&\cr
height2.0pt& \omit&& \omit&& \omit&& \omit&& \omit&\cr
&$ \hfill \omega \hfill$&
&$ \hfill 0.812\pm0.016  \hfill$&
&$ \hfill 0.799\pm0.011\hfill$&
&$ \hfill 0.789\pm 0.011 \hfill$&
&$\hfill 0.782\pm0.0013 \hfill$&\cr
height2.0pt& \omit&& \omit&& \omit&& \omit&& \omit&\cr
&$ \hfill \theta=\omega\nu \hfill$&
&$ \hfill 0.478\pm0.010  \hfill$&
&$ \hfill 0.504\pm0.008\hfill$&
&$ \hfill 0.529\pm0.009  \hfill$&
&$\hfill 0.553\pm0.012 \hfill$&
\cr
height2.0pt& \omit&& \omit&& \omit&& \omit&& \omit&\cr
\fileth }}$$
\endinsert
\midinsert
$$ \vbox{\elevenpoint\offinterlineskip\tabskip=0pt\halign to \hsize
{& \vrule#\tabskip=0em plus1em & \strut\ # \
& \vrule#& \strut #
& \vrule#& \strut # 
& \vrule#& \strut # 
& \vrule#& \strut # 
&\vrule#\tabskip=0pt\cr
\noalign{\centerline{Table \tteps}\tableskip}
\noalign{\centerline{\it
Critical exponents of the
$O(N)$ models from $\varepsilon$-expansion (present work).}
\tableskip}
\fileth
height2.0pt& \omit&& \omit&& \omit&&\omit&& \omit&\cr
&$ \hfill N \hfill$&&$ \hfill 0
\hfill$&&$ \hfill 1 \hfill$&&$ \hfill 2
\hfill$&&$ \hfill 3 \hfill$&\cr
height2.0pt& \omit&& \omit&& \omit&& \omit&& \omit&\cr
\fileth
height2.0pt& \omit&& \omit&& \omit&& \omit&& \omit&\cr
&
$ \hfill\textstyle \gamma\ (\free)\atop \textstyle\gamma\ (\bc) \hfill$&&
$ \hfill\textstyle 1.1575\pm0.0060 \atop \textstyle 1.1571\pm0.0030\hfill$&&
$ \hfill\textstyle 1.2355\pm 0.0050\atop \textstyle 1.2380\pm0.0050\hfill$&&
$ \hfill\textstyle 1.3110\pm0.0070\atop \textstyle 1.317 \hfill$&&
$\hfill \textstyle 1.3820\pm0.0090\atop\textstyle 1.392 \hfill$&\cr
height2.0pt& \omit&& \omit&& \omit&& \omit&& \omit&\cr
&$ \hfill \textstyle \nu\ (\free)\atop \textstyle\nu\ (\bc) \hfill$&&
$ \hfill\textstyle 0.5875\pm0.0025 \atop \textstyle 0.5878\pm0.0011\hfill$&&
$ \hfill\textstyle 0.6290\pm0.0025\atop\textstyle 0.6305\pm0.0025\hfill$&&
$ \hfill \textstyle 0.6680\pm0.0035 \atop \textstyle 0.671\hfill$&&
$\hfill\textstyle 0.7045\pm0.0055 \atop\textstyle 0.708\hfill$&\cr
height2.0pt& \omit&& \omit&& \omit&& \omit&&\omit&\cr
&$ \hfill \textstyle \eta\ (\free)\atop \textstyle\eta\ (\bc) \hfill$&&
$ \hfill\textstyle 0.0300\pm0.0050\atop\textstyle 0.0315\pm0.0035 \hfill$&&
$ \hfill\textstyle 0.0360\pm0.0050\atop\textstyle 0.0365\pm0.0050\hfill$&&
$ \hfill\textstyle 0.0380\pm0.0050 \atop\textstyle 0.0370 \hfill$&&
$\hfill\textstyle 0.0375\pm0.0045\atop\textstyle 0.0355 \hfill$&\cr
height2.0pt& \omit&& \omit&& \omit&& \omit&& \omit&\cr
&$ \hfill \textstyle \beta\ (\free)\atop \textstyle\beta\ (\bc) \hfill$&&
$ \hfill\textstyle 0.3025\pm0.0025\atop\textstyle 0.3032\pm0.0014 \hfill$&&
$ \hfill\textstyle 0.3257\pm0.0025\atop\textstyle 0.3265\pm0.0015\hfill$&&
$ \hfill 0.3465\pm0.0035  \hfill$&&
$\hfill 0.3655\pm0.0035 \hfill$&\cr
height2.0pt& \omit&& \omit&& \omit&& \omit&& \omit&\cr
&$ \hfill \omega \hfill$&&
$ \hfill 0.828\pm0.023  \hfill$&&
$ \hfill 0.814\pm0.018 \hfill$&&
$ \hfill 0.802\pm 0.018 \hfill$&&
$\hfill 0.794 \pm0.018 \hfill$&\cr
height2.0pt& \omit&& \omit&& \omit&& \omit&& \omit&\cr
&$ \hfill \theta \hfill$&&
$ \hfill 0.486\pm0.016  \hfill$&&
$ \hfill 0.512\pm0.013 \hfill$&&
$ \hfill 0.536\pm 0.015 \hfill$&&
$\hfill 0.559 \pm0.017 \hfill$&\cr
height2.0pt& \omit&& \omit&& \omit&& \omit&& \omit&\cr
\fileth}}$$
\endinsert

 \midinsert
$$ \vbox{\elevenpoint\offinterlineskip\tabskip=0pt\halign to \hsize
{& \vrule#\tabskip=0em plus1em & \strut\ # \
& \vrule#& \strut #
& \vrule#& \strut # 
&\vrule#\tabskip=0pt\cr
\noalign{ \centerline{Table \ttn4}\tableskip}
\noalign{\centerline{\it  Critical exponents in the
$O(4)$ models from $d=3$ and $\varepsilon$-expansion (present work).} \tableskip}
\fileth
height2.0pt& \omit&& \omit&& \omit&\cr
&$ \hfill \hfill$&&$ \hfill d=3\hfill$&&$ \hfill \varepsilon:\free,\bc
\hfill$&\cr 
height2.0pt& \omit&& \omit&& \omit&\cr
\fileth
height2.0pt& \omit&& \omit&& \omit&\cr
&$ \hfill \g^*_{\rm Ni} \hfill$
&&$ \hfill 1.377\pm 0.005  \hfill$
&&$ \hfill\hfill$&\cr
height2.0pt& \omit&& \omit&& \omit&\cr
&$ \hfill g^* \hfill$
&&$ \hfill 17.30\pm 0.06  \hfill$
&&$ \hfill \hfill$
&\cr
height2.0pt& \omit&& \omit&& \omit&\cr
&$ \hfill \gamma \hfill$
&&$ \hfill 1.456\pm 0.010 \hfill$
&&$ \hfill 1.448\pm 0.015\ ,\ 1.460\hfill$&
\cr
height2.0pt& \omit&& \omit&& \omit&\cr
&$ \hfill \nu \hfill$
&&$ \hfill 0.741\pm 0.006  \hfill$
&&$ \hfill 0.737\pm 0.008\ , \ 0.742 \hfill$&
\cr
height2.0pt& \omit&& \omit&& \omit&\cr
&$ \hfill \eta \hfill$
&&$ \hfill 0.0350\pm0.0045\hfill$
&&$ \hfill 0.036\pm 0.004\ , \ \ 0.033 \hfill$&
\cr
height2.0pt& \omit&& \omit&& \omit&\cr
&$ \hfill \beta \hfill$
&&$ \hfill 0.3830\pm 0.0045\hfill$
&&$ \hfill 0.3820\pm 0.0025\hfill$
&\cr
height2.0pt& \omit&& \omit&& \omit&\cr
&$ \hfill \alpha \hfill$
&&$ \hfill -0.223\pm 0.018 \hfill$
&&$ \hfill -0.211\pm 0.024 \hfill$&\cr
height2.0pt& \omit&& \omit&& \omit&\cr
&$ \hfill \omega \hfill$
&&$ \hfill 0.774\pm 0.020  \hfill$
&&$ \hfill 0.795\pm 0.030\hfill$
&\cr
height2.0pt& \omit&& \omit&& \omit&\cr
&$ \hfill \theta \hfill$&
&$ \hfill0.574\pm 0.020\hfill$&
&$ \hfill 0.586\pm 0.028\hfill$&\cr
height2.0pt& \omit&& \omit&& \omit&\cr
\fileth }}$$
\endinsert

\section Renormalized $\phi\sp{4}$ field theory: $\varepsilon$-expansion and
3D perturbation series

In this article the general framework is 
the $(\phib^2)^2$,  $O(N)$ symmetric, quantum field theory whose 
bare action is:\sslbl\ssefact
$${\cal H} (\phib)  = \int \left\{ \ud \left[
\partial_\mu \phib(x)\right]^2+\ud \lambda_2 \phib^2(x) +\frac{1}{4!}\lambda_4
[\phib^2(x)]^2 \right\} \d^{d}x\,.  \eqnd\eaction $$
We recall that near the critical temperature $T_c$  $\lambda_2$ is a linear
measure of the temperature. If we denote 
by $\lambda_{2c}$ the value for which the theory becomes massless ($T=T_c$)
then the parameter $t$
$$t=\lambda_2-\lambda_{2c}\propto T-T_c \,,\eqnd\edeftemp $$
characterizes the deviation from the critical temperature.\par

The $(\phib^2)^2$ field theory is renormalizable in four dimensions, and to
eliminate UV divergences (for $d<4$ the theory is super-renormalizable) 
one introduces renormalized correlation functions. 
This involves choosing a renormalization scheme and then trading the
bare parameters $\lambda_2,\lambda_4$ for a (scheme-dependent)
renormalized mass $m$ and dimensionless coupling $g$. The mass parameter $m$
is proportional to the physical mass, or inverse correlation length, of the
high temperature phase. It behaves for $t\propto T-T_c\to 0_+$ as $m\propto
t^\nu$, where $\nu$ is the correlation length exponent (see \rbook~for
details). 
\par
Renormalization group (RG) arguments tell us that the long distance
properties of the massless (critical) theory are governed by non-trivial
IR fixed points $g^*$, solution of the equation 
$$\beta(g^*)=0\,, \quad{\rm with }\ \beta'(g^*)=\omega>0\,.$$  
\par
The anomalous dimensions $\eta (g)$ and  $\eta_2 (g)$, of the renormalized
field $\phib_\r =\phib/\sqrt{Z}$ and of the 
renormalized composite operator $[\phib^2]_\r=(Z_2/Z)\phib^2$ respectively,
evaluated at $g=g^*$ then yield the two independent combinations of critical
exponents (e.g.~$\eta=\eta (g^*)$).  
The explicit forms of the RG functions $\beta(g), \eta (g), \eta_2(g)$ depend
on the specific renormalization scheme. 
\par

The space dimension relevant for statistical physics is $d=3$ (occasionally
$d=2$). In this case one faces a serious problem:  ordinary perturbative
expansion in 
$g$ in the massless theory is IR divergent for any fixed dimension $d$, $d<4$.
A solution to this problem was first provided by Wilson--Fisher's
$\varepsilon=4-d$-expansion. The idea is to avoid IR problems by 
expanding in $\varepsilon=4-d$ as well as in the coupling constant $g$.
IR singularities are then only logarithmic and can be dealt with.
The expansion to the highest order presently available have been performed
within the minimal subtraction $\MS$ scheme, \cite{\repsilonv,\repsigood}. 
In this scheme the $d$-dimensional RG beta function takes
the exact form 
$$\beta_{(\MS)} (g_{\MS}, \epsilon)=-\varepsilon g_{\MS} + f(g_{\MS})
=-\varepsilon g_{\MS}+O(g_{\MS}^2).$$
The fixed point equation 
$$\beta_{(\MS)} (g_{\MS}^*, \epsilon)=0\,,$$
can be solved in the form of an $\varepsilon$-expansion. The
$L$-loop expansion of the $\beta$-function then yields 
$g^*_{\MS}$ up to order $\varepsilon^{L}$. 
By replacing 
$g_{\MS}$ by  $g^*_{\MS}$
in the perturbative  expansion of
the  anomalous dimensions $\eta, \eta_2$ 
one finally obtains  the $\varepsilon$-expansion of critical exponents. 
Note that while $g^*$ is scheme-dependent,
the $\varepsilon$-expansion for universal quantities is scheme-independent. 
\par

While this method directly yields a formal expansion for exponents 
a practical problem arises when one wants to determine exponents for a
physical value of $\varepsilon$ like $\varepsilon=1$ ($d=3$).
Indeed the $\varepsilon$-expansion is divergent as has been first empirically
noted in \rBLGZJN~and later confirmed by the large order behaviour analysis.
A summation method is therefore required to obtain accurate results.
\par
Following Parisi's suggestion \rParisi~perturbation series have also been
calculated directly in three dimensions in the framework of the massive
renormalized theory where  correlation functions $\Gamma^{(n)}_\r$ of the
renormalized field $\phib_\r$ are fixed by the normalization conditions 
\eqna\egrzerom
$$ \eqalignno{ \Gamma^{(2)}_\r (p;m ,g) & = m^2 +p^2 + O \left(p^4 \right), &
\egrzerom{a} \cr \Gamma^{(4)}_\r \left(p_i=0;m ,g \right) & = m g\,. &
\egrzerom{b}  \cr} $$
One may be surprised by the introduction of coupling and field
renormalizations in a super-renormalizable theory. The reasons are simple,
the bare coupling constant becomes infinite when the physical mass goes to
zero. Simultaneously the field renormalization also diverges (see \cite{\rbook,\reviewLGZJ}).
\par
Series up to six loops obtained in this scheme in Ref.~\cite{\rNMB} for
$N=0\cdots3$ have been generalized in Ref. \cite{\rAnSo} to any $N$. 
Only recently in \cite{\rBGNMU} the results for $\eta$ and $\eta_2$
(but not for $\beta(g)$) have been extended to seven loops for
 $N=0\cdots3$ (see Appendix \ssapp). One problem here is that 
the value  $g^*$ of the fixed point coupling is affected by summation errors
on the $\beta$-function. Errors on $g^*$ then induces systematic
errors for all critical exponents (see section \ssUQsum). 

\section Series summation 

Perturbative quantum field theory generates divergent series. Summing such
series by simply adding successive terms is  meaningful only as long as
coupling constants remain small enough (like in QED). Here however the
expansion parameter, the fixed point value $g^*$, is a number of order 1: one
therefore faces the problem of evaluating the sum of divergent series in a
non-trivial regime. \sslbl\ssUQsum \par
In this article the Borel--Leroy transformation has been used, followed by a
conformal mapping \refs{\rLoef} (a new version of the method developed in
\rLGZJiii~for critical exponents) to sum the series. We recall that
the Borel summability of the $\phi^4$ theory in two and three dimensions
has been established in \refs{\rBorsom}.
\par
Let ${\cal S}(z)$ be any (Borel summable)
function whose series has to be summed. We transform the
series:
$${\cal S}(z)=\sum_{k=0} {\cal S}_k z^k, \eqnd\eseries $$
into:
$$ {\cal S}(z)= \sum_{k=0}^{}B_{k}(b)
\int^{\infty}_{0}t^{b}\e^{-t}u^k(zt) \d t\,,\eqnd\eBoreltr $$
with: 
$$ u(s)={ \sqrt{1+as}- 1\over \sqrt{1+as}+ 1} \,. \eqnn $$ 
The coefficients $B_k$ are calculated by expanding in powers of $z$ the
r.h.s.~of equation \eBoreltr~and identifying with expansion \eseries. 
The constant $a$ has been determined by the large order behaviour analysis.
The explicit values are
$$a=0.147774232 \times {9\over N+8}\ , \eqnd\eaLOB $$ 
for the perturbative expansion in $d=3$ dimensions
and
$$a= {3\over N+8}\ , \eqnn $$ 
for the $\varepsilon=4-d$ expansion.
We map the Borel plane, cut at the instanton singularity $s=-1/a$,
onto a circle in the $u$-plane in such a way to enforce
maximal analyticity and thus to optimize the rate of convergence 
(for details see e.g.~\rbook).
\medskip
{\it Additional technical details.} 
Following an idea introduced in \rLGZJiii~for the summation
of the $\varepsilon$-expansion we have in addition made a
homographic transformation on the coupling constant $z$ to displace 
possible singularities in the complex $z$-plane:
\def\param{q}
$$z=z'/(1+ \param z'). \eqnn $$
We have looked for values of the parameters $\param$ and $b$
for which the results were specially insensitive to the order $k$:
in practice the absolute differences of results
corresponding to three successive orders have been minimized. When several
solutions were found 
the less sensitive solution was chosen. Moreover the value of $b$ had to
stay within a reasonable range around the value predicted by the large order
behaviour.  
\par
For each series ${\cal S}(z)$ we have applied the summation procedure
both to $\cal S$ and $1/{\cal S}$. Finally we have introduced ``shifts",
for each series summing 
$${\cal S}_s(z)=\left({\cal S}(z)-\sum_{k=0}^{s-1}{\cal S}_k z^k\right)/z^s,$$
In practice only the cases $s=0$ (no subtractions) and $s=1,2$ have proven
useful. Thus for each exponent we have obtained six results whose
spread gives an indication of summation errors.
\par
In some examples (in particular $g^*$) shifts have produced strongly
oscillating results. It has appeared that it would be useful to somehow
interpolate between shifted series. An idea, new to this work,
has been to consider the combination 
$${\cal S'} \equiv (1+\z g){\cal S}\,,\eqnd\eerre $$ 
where $\z$ has been used as a third variational parameter
(dividing of course the final result by the factor $(1+\z g)$).
The precise value of $r$ has been obtained by minimizing the dependence in
$b$.
This new additional parameter has proven quite useful: it has allowed, as
expected, to obtain series with better apparent convergence as well as better
general consistency. It has also revealed that in a few cases the apparent
convergence at $r=0$ was deceptive, the results being unstable 
with respect to a variation of $\z$. These cases had already been singled out 
by the extreme values of the optimal $b,\param$ parameters.\par
The main consequence of this new approach has been a decrease in the values of
$g^*$ though the length of the series has not changed
(better agreement between different shifts 
at nonzero $r$, see Fig.1), and of $\gamma$ for $N=0$ and $d=3$ (best $r=0$
values of the exponent revealed to be unstable). 
\medskip
{\it Errors.} The summation error for any quantity ${\cal S}(z)$ 
has been estimated  by looking at differences between successive orders,
sensitivity to the 
parameters and spread between all results concerning
the same exponent (this has also involved checking scaling relations).  
\par
In the case of the 3D perturbative expansion, the total error
for each exponent ${\cal S}$ is the sum of the intrinsic summation error at 
fixed $\g^*$, $\Delta{\cal S}$, and the error induced by the error in $\g^*$,
$\Delta\g^*$: 
$${\cal S} ={\cal S}^* \pm  \Delta{\cal S}  
\pm  \left({d {\cal S}\over d\g}\right)_{\g^*} \Delta\g^* .\eqnn $$
We thus also give the derivatives of exponents with respect to $\g^*$.
Two derivatives are displayed in Table \ttdg, all other ones can be deduced
from scaling relations. The reader can thus infer the sensitivity of exponents
to a change in the values and errors of $g^*$.
\midinsert
$$ \vbox{\elevenpoint\offinterlineskip\tabskip=0pt\halign to \hsize
{& \vrule#\tabskip=0em plus1em & \strut\ # \ 
& \vrule#& \strut #
& \vrule#& \strut #  
& \vrule#& \strut #  
& \vrule#& \strut #  
& \vrule#& \strut #  
&\vrule#\tabskip=0pt\cr
\noalign{\centerline{Table \ttdg} \tableskip}
\noalign{\centerline
{\it Critical exponents: Sensitivity to $\g^*$ determination.}
\tableskip} \fileth
height2.0pt& \omit&& \omit&& \omit&& \omit&&\omit&& \omit&\cr
&$\hfill N \hfill$&&$ \hfill 0\hfill$&&$ \hfill 1 \hfill$&&$ \hfill 2 \hfill$&
&$ \hfill 3 \hfill$&&$ \hfill 4 \hfill $&\cr
height2.0pt& 
\omit&&\omit&&\omit&&  \omit&& \omit&&\omit&\cr \fileth
height2.0pt& \omit&&\omit&&\omit&& \omit&& \omit&& \omit&\cr 
&$\hfill \d \gamma/ \d\g^* \hfill$&
&$ \hfill 0.10 \hfill$&
&$ \hfill 0.18\hfill$&
&$ \hfill 0.28\hfill$&
&$ \hfill 0.39\hfill$&
&$\hfill  0.50\hfill$&\cr 
&$\hfill \d\nu / \d\g^* \hfill$&
&$ \hfill 0.069 \hfill$& 
&$ \hfill 0.11\hfill$&
&$ \hfill 0.17\hfill$&
&$ \hfill 0.22\hfill$&&$
\hfill  0.29\hfill$&\cr 
height2.0pt&\omit&&\omit&& \omit&& \omit&& \omit&& \omit&\cr
\fileth}}$$
\endinsert
Let us stress here that we quote in our tables the total combined error  
(as well in \cite{\rLGZJi,\rLGZJii}) while only the intrinsic summation error
is reported in the Table \ttnic for the alternative result of \cite{\rBGNMU}.
\par
For what concerns the $\varepsilon$-expansion the
total error is directly given by the intrinsic summation error
of each exponent and the situation is in principle more favourable:
the only problem then is that the available series
are shorter (they are technically more difficult to obtain)
and the summation error is then bigger!
\medskip
{\it Remarks.} The comparison between results coming from direct $d=3$ series
and $\varepsilon$-expansion is useful not only to test the accuracy of our
numerical methods. Their consistency is also important to test various
assumptions or properties.\par
In the case of the $d=3$ expansion we assume more analyticity in the Borel
plane as has been rigorously proven. Semiclassical instanton analysis
indicates that our 
assumption is quite plausible but this is not a proof. Moreover several
authors (see e.g. \cite{\rNickel,\rPelVic}) have argued that RG functions are not regular at $g=g^*$. We have of
course checked that these singularities, if they exist, are weak. Numerical
evidence is that all RG functions are at least differentiable 
at $g=g^*$ (including
$\beta'(g)$ which yields $\omega$). We cannot of course exclude the situation
where these singularities are so weak as to escape detection, but strong
enough to influence results at the level of accuracy at which exponents are
determined. Our apparent errors could then be underestimated.
Nevertheless it should be emphasized that if the hypothesis of analyticity 
in the cut Borel plane holds, the Borel summation should anyway converge
asymptotically even in presence of confluent singularities. 
\nref\ruvren{G. Parisi, Carg\`ese Lectures 1977, Vol B39 (Plenum New York
1979).} 
\par
For what concerns the $\varepsilon$-expansion problems are more serious, since
Borel summability has not even be proven. Moreover there are indications that
{\it UV-renormalon}\/ singularities could prevent Borel summability,
\cite{\ruvren}. These
singularities are related to the large momentum behaviour of renormalized
perturbation theory (the ``Landau ghost" problem). 
A plausible conjecture is that quantities related to the massless theory are 
renormalon-free since they can be calculated in the theory with UV cut-off.
This in particular applies to critical exponents. Instead the question remains
open for quantities only defined in the massive renormalized theory, like the
fixed point coupling constant $g^*(\varepsilon)$ defined by
\egrzerom. Note that, because the $\varepsilon$ series are rather short,
empirical evidence is weak.
\section  Pseudo-epsilon expansion and exponents' correlation analysis

In \cite{\rLGZJii}, a method was introduced to try to circumvent the problem
of systematic errors induced by an error in the determination of $g^*$: the
so-called 
pseudo-epsilon expansion. The idea is to mimic the $\varepsilon$-expansion and
introduce a new parameter $\pseudoe$ in terms of which $\g^*$ is expanded
as well as all critical exponents. \sslbl\sssexp
\par
The $d=3$ beta-function in the scheme
Eqs.~\egrzerom{} has the form:
$$ \beta(g)= -g+\beta_2(g) $$
where $\beta_2$ begins at order $g^2$ with a positive coefficient of order 1.
We then replace the $\beta$-function by a new function $\beta(g,\pseudoe)$
$$\beta(g,\pseudoe)=-\pseudoe g +\beta_2(g), $$
and expand  $g^*(\pseudoe)$, the solution to $\beta(g,\pseudoe)=0$,
in powers of $\pseudoe$. Eventually we have to sum the series
for the value of $\pseudoe=1$ to recover the initial equation.\par
This method has been systematically used in {\rLGZJii}, and this explains why
some of the new values of exponents we obtain in this work differ less from
the previous values of {\rLGZJii} than the change in $g^*$ would lead to
expect.\par   
To apply the same method here, we face the problem that the series for the
$\beta$-function have not been extended to seven loops, and therefore for the
exponents $\gamma$ or $\nu$, for example, the information of the additional
seven loops term cannot
be used. However since $\eta(g)$ starts only  at order $g^2$,
to determine $\eta$ at loop order $L$, $g^*(\pseudoe)$ is required  only at
loop order $L-1$. This also applies to the exponent $\delta$ which only
depends on $\eta$, and to which we have equally applied the summation
procedures.\par 
It follows that for $N=0,1,2,3$  the pseudo-epsilon
expansion yields genuine seven loop information on $\eta, \delta$,
with apparent errors much smaller compared to six-loop results.
\par 
Still to try to circumvent the problem of shorter $\beta(g)$ series and 
$g^*$ determination, we have in this work introduced another idea.
We have directly eliminated the coupling constant between a pair of
independent exponents. For example we have inverted the relation
$g\mapsto 2-1/\nu$
$$g(\nu)=\sum g_k(2-1/\nu)^k, $$
and then expressed other exponents as series in $2-1/\nu$. In this way we
have obtained correlation curves between exponents, which all eventually can
be translated into relations $\eta(\nu)$. We have applied the same idea
starting from the exponents $\gamma,\beta$, expanding in powers of 
$1-1/\gamma$ and $4-1/\beta$.
\par
With this in mind it is interesting to consider the derivatives  ${\d\eta /
\d\nu}$ at the fixed point, which we thus display in Table \ttds. Other
derivatives can be deduced, using scaling relations. The correlation line can
be fixed by taking a point from the list of Table \ttd3.\par
Finally let us note that we can push this idea up to expanding the RG
$\beta$-function in powers of for example $2-1/\nu$ and solving directly the 
fixed point equation $\beta(\nu^*)=0$. \par
We have tried this idea but the main problem we have faced is that the
general structure of series generated by this set of transformations is rather
complicated and therefore the apparent errors are quite large (a problem which
already limits the accuracy of the pseudo-epsilon expansion). Therefore the
method has mainly be used as a check of consistency among the data generated
by more direct summation. It is possible that more accurate constraints
could be obtained with more work to better understand the convergence of these
new series, but we have eventually generated so many data that it became
difficult to analyze all of them with the same care.
\midinsert
$$ \vbox{\elevenpoint\offinterlineskip\tabskip=0pt\halign to \hsize
{& \vrule#\tabskip=0em plus1em & \strut\ # \ 
& \vrule#& \strut #
& \vrule#& \strut #  
& \vrule#& \strut #  
& \vrule#& \strut #  
& \vrule#& \strut #  
&\vrule#\tabskip=0pt\cr
\noalign{\centerline{Table \ttds} \tableskip}
\noalign{\centerline
{\it Critical exponents: Correlation between exponents at the fixed point.}
\tableskip} \fileth
height2.0pt& \omit&& \omit&& \omit&& \omit&&\omit&& \omit&\cr
&$\hfill N \hfill$&&$ \hfill 0\hfill$&&$ \hfill 1 \hfill$&&$ \hfill 2 \hfill$&
&$ \hfill 3 \hfill$&&$ \hfill 4 \hfill $&\cr
height2.0pt& 
\omit&&\omit&&\omit&&  \omit&& \omit&&\omit&\cr \fileth
height2.0pt& \omit&&\omit&&\omit&& \omit&& \omit&& \omit&\cr 
&$\hfill {\d\eta/\d\nu}~~~~~ (d=3) \hfill$&
&$ \hfill 0 .83\hfill$& 
&$ \hfill 0.59\hfill$&
&$ \hfill 0.43\hfill$&
&$ \hfill 0.32\hfill$&
&$\hfill  0.27\hfill$&
\cr 
height2.0pt&\omit&&\omit&& \omit&& \omit&& \omit&& \omit&\cr
\fileth}}$$
\endinsert

\section Numerical results

Let us first consider $d=3$ results. The values of $g^*$ have been obtained by
looking for the zeros of the summed RG function $\beta (g) $.
The various methods explained in section \ssUQsum~have been used, shifts
$0,1,2$, generating three set of values for each $N$, depending on three
parameters 
$b,\param,\z$. Quoted errors for $g^*$ reflect the apparent convergence with
the order $k$ ($k\le 7$), the sensitivity of $g^*$ to a variation of the
parameters $b,\param,\z$ around optimal values as well as the spread between
different shifts. 
As additional checks we have looked for the zeros of the function
$\nu(g)\beta(g)$ (a rather arbitrary choice with the weak motivation that the
derivative yields the exponent $\theta=\omega\nu$) and calculated $g^*$ from
the pseudo-expansion (see \cite{\rLGZJii} for details). 
Final results are reported in Table \ttd3 (Table \ttn4 for $N=4$).
\sslbl\ssnumres 
%
%
\par
For what concerns
the values of exponents,
 we have summed the seven (six for $N=4$) loop series 
at the values of $g^*$ determined before. We have summed independently
the five exponents $\gamma,\beta,\nu,\delta,\eta$ by using
the three parameters $b,\param,\z$ for each exponent (and its inverse) 
and  shifts $0,1$ (shift $2$ was considered only as a check).
Again, errors have been estimated by decreasing the order and looking at the
spread between  summation of different equivalent  series,  
as explained in section \ssUQsum.
Additional checks  have been derived from pseudo-expansion and
exponents' correlation analysis, introduced in section \sssexp. 
Table \ttd3 and Table \ttn4 report the results of the analysis.

For what concerns the $\varepsilon$-expansion the procedure is the same as for
the $d=3$ series, apart from the fact that the $g^*$ step is bypassed, the
series being summed at $\varepsilon =1$ for the physical dimension three.

More precisely we have summed the genuine $\varepsilon$ series for the
exponents  
(called ``free" in Table \tteps ) and we summed as well the modified series
$$ {\tilde {\cal S}}(d)\equiv {{ {\cal S}}(d) 
- { {\cal S }(2)}\over d-2} $$
in which for each exponent is imposed the exact value at  $d=2$ (referred to
"bc", with boundary conditions, in Table \tteps).
For $N=0,1$ the $d=2$ exact exponents are obtained from the underlying 
conformal theories,
for $N=2$ from the identification with the Kosterlitz--Thouless
transition, while for $N>2$ the behaviour near 
$d=2$  can be obtained from the  $O(N)$ non-linear
$\sigma$-model.  
 
The general conclusions are the following: by imposing boundary conditions,
we decrease the apparent errors for $N=0$. For $N=1$ apparent errors remain 
about the same but the central values are slightly modified. 
For $N\ge 2$ the convergence of the series with boundary condition is worse
and we report in Table \tteps~only central values for the exponents for which
the convergence seems reasonable. Errors can approximately been inferred
from the difference with the free values and the corresponding apparent
errors. 

\section Updated values for the $N=1$ equation of state and critical exponents

The new values of the critical exponents obtained in this work
directly affect the determination of the scaling
equation of state for the $N=1$, $d=3$ case, by the method presented in
\cite{\rRGZJ}. We thus report here the new estimates. 
The results for the $\varepsilon$-expansion have been revised too
(in particular the errors on amplitude ratios have been
reconsidered).  

We recall that our starting point was
an estimation of the values of coefficients $F_k$ of small magnetization
expansion for the derivative of the effective potential $V$ (free energy)
with respect to the scaled renormalized field $z$ (magnetization)
$${\partial V\over \partial z}=z+\frac{1}{6}z^3+\sum_{k=2}F_{2k+1}(g)
z^{2k+1}\,.\eqnn $$  
These coefficients $F_k$ have been summed in \cite{\rRGZJ} by using the
available series up to five loops \refs{\rBBMN,\rHaDo,\rRAJA}
and are reported in Table \tteqst, compared with results
of other techniques (a misprint in the last digit of the value of $\g^*$
has been corrected).
\sslbl\ssupdate
A uniform approximation for the equation of state has then be provided
by the determination of the auxiliary 
function $h(\theta)$ defined by the reparametrization
(see also \cite{\rJoseph,\rWZAN,\rNA}):
 $$\eqalignno{
z&=\rho \theta/ (1-\theta^2)^\beta& \eqnd\efase\cr
h(\theta)&=\rho^{-1}
\left(1-\theta^2\right)^{\beta\delta}F\left(z(\theta)\right).
&\eqnd\emagfparii\cr }$$ 
The Order Dependent Mapping technique
 \ref\rODM{R. 
Seznec and J. Zinn-Justin, {\it J. Math. Phys.} 20 (1979) 1398\semi 
J.C. Le Guillou and J. Zinn-Justin, {\it Ann. Phys. (NY)} 147 (1983) 57\semi
R. Guida, K. Konishi and H. Suzuki, {\it Ann. Phys. (NY)} 241 (1995) 152; 249
(1996) 109. For related work see: H. Kleinert and W. Janke, {\it Phys.Lett.} A
206 (1995) 283}. 
 has been used to improve convergence of
the small $\theta$-expansion by an optimal choice of the parameter $\rho$.

The new result coming from the revised values of $\gamma, \beta$ is
$$h(\theta)/\theta=1-0.762(3)\;\theta^2+0.0082(10)\;\theta^4,
\eqnd\mainresult$$
that is obtained from 
$\rho^2=2.86$. 
This expression of $h(\theta)$ has a zero at 
$$\theta_0^2=1.33\ , \eqnn$$
to which corresponds the value of the complex root $z_0$ of $F(z)$,
$|z_0|=2.80$ 
(the phase, given by Eq.~\efase, is $-i\pi \beta$).
\medskip
The revised $\varepsilon$-expansion estimations
of $F_k$ (in Table \tteqst)
and of critical amplitudes presented in this paper are obtained using the revised $\gamma, \beta$ of Table \tteps 
and the following expression of $h (\theta)$ (summed at $\varepsilon=1$):
$$h(\theta)/\theta=1-0.72(6)\;\theta^2+0.0136(20)\;\theta^4.
\eqnd\epsiresult$$
It should be emphatized that 
while critical amplitudes and the equation of state are
universal quantities,
$h(\theta)$ is not a universal function;
in particular the variable $\theta$ of $\varepsilon$-expansion
should not be identified to the corresponding variable of the $d=3$
analysis, because they are defined from a different mapping
Eq.~\efase,
(different $\rho$ and $\beta$). 
It follows that $h(\theta)$ of the two methods (and their errors)
cannot be compared directly.  
\par
Our  value of $g^*$ from $\varepsilon$-expansion
(in Table \tteqst) has been obtained
from our own analysis of $O(\varepsilon^4)$ series of \cite{\rPelVic}.
We report in Table \tteqst also the recent results of \cite{\rPelVicii},
obtained by a direct summation of  $O(\varepsilon^3)$ series for 
$F_k$ (improved by imposing boundary conditions at smaller dimensions).


\medskip
Widom's scaling function $f(x)$ (with $f(-1)=0$ and $f(0)=1$) can easily be
derived by (numerically) solving the following system:
$$
\left\lbrace\eqalign{
&f(x)=\theta^{-\delta} h(\theta)/h(1)\cr 
&x=  \left(-{1-\theta^2\over 1-\theta_0^2 }\right) 
\left( {\theta_0\over \theta} \right)^{1/\beta} 
}\right. \eqnd\eWidomnum 
$$

%
\midinsert
$$ \vbox{\elevenpoint\offinterlineskip\tabskip=0pt\halign to \hsize
{& \vrule#\tabskip=0em plus1em & \strut\ # \ 
& \vrule#& \strut #
& \vrule#& \strut #  
& \vrule#& \strut #  
& \vrule#& \strut #  
&\vrule#\tabskip=0pt\cr
\noalign{\centerline{Table \tteqst} \tableskip}
\noalign{\centerline{\it Equation of state.}
\tableskip}
\fileth
height2.0pt& \omit&& \omit&& \omit&&\omit&& \omit&\cr
&$ \hfill  \hfill$&&$ \hfill g^* \hfill$&&$ \hfill
F_5  \hfill$&&$ \hfill F_7 \times 10^4
\hfill$&&$ \hfill F_9 \times 10^5\hfill$&\cr 
height2.0pt& \omit&& \omit&& \omit&& \omit&& \omit&\cr
\fileth
height2.0pt& \omit&& \omit&& \omit&& \omit&& \omit&\cr
&$ \hfill \varepsilon{-\rm exp.}, \hbox{this work} \hfill$
&&$ \hfill 23.3\hfill$
&&$ \hfill 0.0177\pm 0.0010 \hfill$
&&$ \hfill 4.8\pm0.6  \hfill$
&&$\hfill -3.3\pm0.3 \hfill$&\cr
height2.0pt& \omit&& \omit&& \omit&& \omit&& \omit&
\cr
&$ \hfill \varepsilon{-\rm exp.}, \cite{\rPelVic,\rPelVicii} \hfill$
&&$ \hfill 23.4 \pm 0.1\hfill$
&&$ \hfill 0.01715\pm 0.00009 \hfill$
&&$ \hfill 4.9\pm0.6  \hfill$
&&$\hfill -5.5\pm 4 \hfill$&\cr
height2.0pt& \omit&& \omit&& \omit&& \omit&& \omit&\cr
&$ \hfill d=3,  \hbox{this work} \hfill$
&&$ \hfill 23.64\pm 0.07  \hfill$
&&$ \hfill0.01711\pm 0.00007 \hfill$
&&$ \hfill 4.9\pm 0.5   \hfill$
&&$\hfill -7\pm5 \hfill$&\cr 
height2.0pt& \omit&& \omit&& \omit&& \omit&&\omit&\cr
&$ \hfill   d=3 ~\rsokolov \hfill$
&&$ \hfill 23.71 \hfill$
&&$ \hfill.01703\hfill$
&&$ \hfill 10 \hfill$
&&$ \hfill  \hfill$&\cr
height2.0pt& \omit&& \omit&& \omit&& \omit&& \omit&\cr
&\hfill   HT~\rreisz \hfill
&&$ \hfill 23.72 \pm 1.49 \hfill$
&&$ \hfill0.0205\pm 0.0052 \hfill$
&&$ \hfill   \hfill$
&&$ \hfill \hfill$&\cr
height2.0pt& \omit&& \omit&& \omit&& \omit&& \omit&\cr
&\hfill  HT~\rZLFish \hfill
&&$ \hfill  24.45\pm0.15\hfill$
&&$ \hfill.017974\pm.00015\hfill$
&&$ \hfill  \hfill$
&&$ \hfill  \hfill$&\cr
height2.0pt& \omit&& \omit&& \omit&& \omit&& \omit&\cr
&\hfill  HT~\rbuco \hfill
&&$ \hfill 23.69 \pm .10 \hfill$
&&$ \hfill.0168\pm 0.0012\hfill$
&&$ \hfill 5.4\pm 0.7 \hfill$
&&$ \hfill -2.3\pm 1.1  \hfill$&\cr
height2.0pt& \omit&& \omit&& \omit&& \omit&& \omit&\cr
&\hfill   MC~\rtsypin \hfill
&&$ \hfill 23.3 \pm 0.5  \hfill$
&&$ \hfill0.0227\pm0.0026 \hfill$
&&$ \hfill  \hfill$
&&$ \hfill  \hfill$&\cr
height2.0pt& \omit&& \omit&& \omit&& \omit&& \omit&\cr
& \hfill MC~\rkimlandau \hfill&&$ 
\hfill 24.5\pm.2 \hfill$
&&$\hfill 0.027\pm0.002 \hfill$
&&$ \hfill 23.6\pm4  \hfill$
&&$ \hfill  \hfill$&\cr
height2.0pt& \omit&& \omit&& \omit&& \omit&& \omit&\cr
& \hfill   ERG~\rwetterich \hfill
&&$ \hfill 28.9  \hfill$&&$ \hfill 0.016 \hfill$
&&$ \hfill 4.3 \hfill$
&&$ \hfill  \hfill$&\cr
height2.0pt& \omit&& \omit&& \omit&& \omit&& \omit&\cr
&\hfill ERG~\rmorris \hfill
&&$ \hfill 20.72\pm 0.01  \hfill$
&&$ \hfill
0.01719  \pm 0.00004 \hfill$
&&$ \hfill 4.9\pm 0.1 \hfill$
&&$ \hfill -5.2\pm 0.3  \hfill$&\cr
height2.0pt& \omit&& \omit&& \omit&& \omit&& \omit&\cr
\fileth}}$$
\endinsert
\medskip
From Eq.~{\eWidomnum} and the revised values of the critical exponents
we can calculate various critical amplitude ratios
that are reported in Tables \ttcramp and \ttcrampb and are
compared with other theoretical and experimental results.
The reader can find all definitions and more details in \cite{\rRGZJ}.
(See also \cite{\rprivman} for a report on the subject)
 
\nref\rCasel{
M. Hasenbusch and K. Pinn, cond-mat/9706003, submitted to J. Phys. A\semi
M. Caselle and M. Hasenbusch, {\it J. Phys.} A30 (1997) 4963
\semi M. Caselle and M. Hasenbusch, Nucl. Phys. Proc. Suppl. 63 (1998) 613.}
\midinsert
$$ \vbox{\elevenpoint\offinterlineskip\tabskip=0pt\halign to \hsize
{& \vrule#\tabskip=0em plus1em & \strut\ # \ 
& \vrule#& \strut #
& \vrule#& \strut #  
& \vrule#& \strut #  
& \vrule#& \strut #  
&\vrule#\tabskip=0pt\cr
\noalign{\centerline{Table \ttcramp} \tableskip}
\noalign{ \centerline {\it Amplitude ratios.}
\tableskip} 
\fileth
height2.0pt& \omit&& \omit&& \omit&&\omit&& \omit&\cr
&$ \hfill $&&$ \hfill A^+/ A^-
\hfill$&&$ \hfill C^+/C^- \hfill$&&$ \hfill
R_c \hfill$&&$ \hfill R_\chi\hfill$&\cr
height2.0pt& \omit&& \omit&& \omit&& \omit&& \omit&\cr
\fileth
height2.0pt& \omit&& \omit&& \omit&& \omit&& \omit&\cr
&$ \hfill \varepsilon-{\rm exp.},\cite{\rbervil,\rNA}
 \hfill$&&$ \hfill 0.524\pm 0.010
\hfill$&&$ \hfill 
4.9 \hfill$&&$ \hfill  \hfill$&&$
\hfill 1.67 \hfill$&\cr
&$ \hfill \varepsilon-{\rm exp.}$, this work \hfill&
&$ \hfill 0.527\pm0.037 \hfill$&
&$ \hfill 4.73\pm 0.16 \hfill$&
&$ \hfill 0.0569\pm0.0035 \hfill$&
&$\hfill 1.648\pm0.036 \hfill$&\cr
height2.0pt& \omit&& \omit&& \omit&&
\omit&& \omit&\cr &$ \hfill d=3,\rBBMN~\hfill$&&$ \hfill 0.541\pm
0.014 \hfill$&&$ \hfill  
4.77\pm 0.30\hfill$&&$ \hfill 0.0594\pm 0.001\hfill$&&$ \hfill 1.7
\hfill$&\cr 
height2.0pt& \omit&& \omit&& \omit&&
\omit&& \omit&\cr &$ \hfill d=3,\hfill$ this work
&&$ \hfill 0.537\pm0.019 \hfill$
&&$ \hfill  4.79\pm 0.10\hfill$
&&$ \hfill 0.0574\pm 0.0020\hfill$
&&$ \hfill 1.669\pm0.018\hfill$&\cr 
height2.0pt& \omit&& \omit&& \omit&& \omit&&\omit&\cr
&$ \hfill \hbox{HT \cite{\rliu,\raharony}} \hfill$&&$ \hfill 0.523\pm 0.009 \hfill$&&$ \hfill
4.95 \pm 0.15 
\hfill$&&$ \hfill 0.0581\pm0.0010 \hfill$&&$ \hfill 1.75
\hfill$&\cr 
height2.0pt& \omit&& \omit&& \omit&& \omit&&\omit&\cr
&$ \hfill \hbox{MC\rCasel} \hfill$&&$ \hfill 0.560\pm0.010 \hfill$&&$ \hfill
4.75 \pm 0.03 
\hfill$&&$ \hfill  \hfill$&&$ \hfill \hfill$&\cr 
height2.0pt& \omit&& \omit&& \omit&& \omit&& \omit&\cr
&$ \hfill {\rm bin.~mix.} \hfill$&&$ \hfill 0.56\pm0.02
\hfill$&&$ \hfill 
4.3\pm 0.3 \hfill$&&$ \hfill 0.050\pm0.015  \hfill$&&$ \hfill  1.75\pm0.30
\hfill$&\cr  
height2.0pt& \omit&& \omit&& \omit&& \omit&& \omit&\cr
&$ \hfill {\rm liqu.-vap.} \hfill$&& \hfill 0.48--0.53 \hfill&& \hfill 
4.8{--}5.2 \hfill&&$ \hfill 0.047\pm0.010  \hfill$&&$ \hfill  1.69\pm0.14
\hfill$&\cr 
height2.0pt& \omit&& \omit&& \omit&& \omit&& \omit&\cr
&$ \hfill {\rm magn.~syst.} \hfill$&& \hfill 0.49{--}0.54 \hfill&&$ \hfill 
4.9\pm 0.5 \hfill$&&$ \hfill   \hfill$&&$ \hfill  
\hfill$&\cr 
height2.0pt& \omit&& \omit&& \omit&& \omit&& \omit&\cr
\fileth}}$$
\endinsert

\midinsert
$$ \vbox{\elevenpoint\offinterlineskip\tabskip=0pt\halign to \hsize
{& \vrule#\tabskip=0em plus1em & \strut\ # \ 
& \vrule#& \strut #
& \vrule#& \strut #  
& \vrule#& \strut #  
&\vrule#\tabskip=0pt\cr
\noalign{\centerline{Table \ttcrampb} \tableskip}
\noalign{\centerline{\it Other amplitude ratios.}\tableskip}
\fileth
%
%
height2.0pt& \omit&& \omit&& \omit&&\omit &\cr
&\omit && $\hfill R_0 \hfill$&&$ \hfill R_3 \hfill$&&$ \hfill {C_4^+/
C_4^-}\hfill$&\cr  
%
%
height2.0pt& \omit  && \omit&& \omit&& \omit&\cr
\fileth
%
height2.0pt& \omit&& \omit&& \omit&& \omit&\cr
&$ \hfill {\rm HT\ series}~\rZLFish\hfill$&&$ \hfill  0.1275\pm0.0003\hfill$&&
$ \hfill 6.4\pm0.2 \hfill$&&$ \hfill -9.0\pm 0.3 \hfill$&\cr
height2.0pt& \omit&& \omit&& \omit&& \omit&\cr
& \hfill$ d=3$, this work\hfill&&$ \hfill  0.12584\pm0.00013\hfill$&&
$ \hfill 6.08\pm0.06 \hfill$&&$ \hfill -9.1\pm 0.6 \hfill$&\cr
& \omit&& \omit&& \omit&&  \omit&\cr
& \hfill $\varepsilon$-expansion, this work \hfill&&$ \hfill
0.127\pm0.002\hfill$&& 
$ \hfill 6.07\pm0.19 \hfill$&&$ \hfill -8.6\pm 1.5 \hfill$&\cr
& \omit&& \omit&& \omit&&  \omit&\cr
\fileth}}$$
\endinsert

\section Conclusions

Before discussing our results, let us review the results for 
critical exponents obtained by other theoretical methods or as well as
experiments.
 
The previous most accurate determinations of the critical exponents of the
$O(N)$ vector model, from quantum field theory and renormalization group,
have been reported in \refs{\rLGZJi,\rLGZJii} and are 
shown in Table \ttlg~(we refer to these results as LG--ZJ). 
In Table \ttnic we report the Murray--Nickel (M--N) predictions
(direct fit of $g$ series)
with the authors' preferred choice of $\g^*$
(they report only summation errors; errors from $\g^*$ should be
added). In Table \ttsok we 
list some values for $N=4$ obtained from Pad\'e Borel summation of
$d=3$ series up to six loops
(see \cite{\rsokolov} where results for many
values of $N>3$ are given).
\nref\rKlodmexp{H. Kleinert, Phys. Rev. D 57 (1998) 2264.}
An analysis  based on Order Dependent Mapping \cite{\rODM} of $d=3$ series
can be found in \cite{\rKlodmexp}.
In Table \ttepsiold we report the previous analysis of $\varepsilon$-expansion
while in Table \ttpelvic we quote some  recent results.   
Other available 
theoretical predictions come from the analysis of High Temperature (HT) series
in lattice models, Table \ttht, and Monte-Carlo (MC) simulations
Table \ttmc. 
Finally in Table \tterg~we report for completeness some estimates from the
truncated ``Exact  Renormalization Group" approach.

\midinsert
$$ \vbox{\elevenpoint\offinterlineskip\tabskip=0pt\halign to \hsize
{& \vrule#\tabskip=0em plus1em & \strut\ # \ 
& \vrule#& \strut #
& \vrule#& \strut #  
& \vrule#& \strut #  
& \vrule#& \strut #  
&\vrule#\tabskip=0pt\cr
\noalign{ \centerline{Table \ttlg}\tableskip}
\noalign{\centerline{\it Estimates of critical exponents in the
$O(N)$ symmetric $(\phib^2)^2_3 $ field theory LG-ZJ.} \tableskip}
\fileth
height2.0pt& \omit&& \omit&& \omit&&\omit&& \omit&\cr
&$ \hfill N \hfill$&&$ \hfill 0
\hfill$&&$ \hfill 1 \hfill$&&$ \hfill 2
\hfill$&&$ \hfill 3 \hfill$&\cr
height2.0pt& \omit&& \omit&& \omit&& \omit&& \omit&\cr
\fileth
height2.0pt& \omit&& \omit&& \omit&& \omit&& \omit&\cr
&$ \hfill \g^* \hfill$&&$ \hfill
1.421\pm0.008  \hfill$&&$ \hfill 1.416\pm0.005
\hfill$&&$ \hfill 1.406\pm 0.004 \hfill$&&$
\hfill 1.391 \pm0.004 \hfill$&\cr
height2.0pt& \omit&& \omit&& \omit&& \omit&& \omit&\cr
&$ \hfill \gamma \hfill$&&$ \hfill 1.1615
\pm0.0020 \hfill$&&$ \hfill 1.2405\pm 0.0015
\hfill$&&$ \hfill 1.316\pm0.0025  \hfill$&&$
\hfill 1.386\pm0.0040 \hfill$&\cr
height2.0pt& \omit&& \omit&& \omit&& \omit&& \omit&\cr
&$ \hfill \nu \hfill$&&$ \hfill 0.5880\pm0.0015
 \hfill$&&$ \hfill 0.6300\pm0.0015
\hfill$&&$ \hfill 0.6695\pm0.0020  \hfill$&&$
\hfill 0.705\pm0.0030 \hfill$&\cr
height2.0pt& \omit&& \omit&& \omit&& \omit&&\omit&\cr
&$ \hfill \eta \hfill$&&$ \hfill 0.027\pm0.004
 \hfill$&&$ \hfill 0.032\pm0.003
\hfill$&&$ \hfill 0.033\pm0.004  \hfill$&&$
\hfill 0.033\pm0.004 \hfill$&\cr
height2.0pt& \omit&& \omit&& \omit&& \omit&& \omit&\cr
&$ \hfill \beta \hfill$&&$ \hfill 0.302\pm0.0015
 \hfill$&&$ \hfill 0.325\pm0.0015
\hfill$&&$ \hfill 0.3455\pm0.0020  \hfill$&&$
\hfill 0.3645\pm0.0025 \hfill$&\cr
height2.0pt& \omit&& \omit&& \omit&& \omit&& \omit&\cr
&$ \hfill \alpha \hfill$&&$ \hfill 0.236\pm0.0045
 \hfill$&&$ \hfill 0.110\pm0.0045
\hfill$&&$ \hfill -0.007\pm0.006  \hfill$&&$
\hfill -0.115\pm0.009 \hfill$&\cr
height2.0pt& \omit&& \omit&& \omit&& \omit&& \omit&\cr
&$ \hfill \omega \hfill$&&$ \hfill
0.80\pm0.04  \hfill$&&$ \hfill 0.79\pm0.03
\hfill$&&$ \hfill 0.78\pm 0.025 \hfill$&&$
\hfill 0.78 \pm0.02 \hfill$&\cr
height2.0pt& \omit&& \omit&& \omit&& \omit&& \omit&\cr
&$ \hfill \theta \hfill$&&$ \hfill
0.470\pm0.025  \hfill$&&$ \hfill 0.498\pm0.020
\hfill$&&$ \hfill 0.522\pm0.0018  \hfill$&&$
\hfill 0.550\pm0.0016 \hfill$&\cr
height2.0pt& \omit&& \omit&& \omit&& \omit&& \omit&\cr
\fileth }}$$
\endinsert

\midinsert
$$ \vbox{\elevenpoint\offinterlineskip\tabskip=0pt\halign to \hsize
{& \vrule#\tabskip=0em plus1em & \strut\ # \ 
& \vrule#& \strut #
& \vrule#& \strut #  
& \vrule#& \strut #  
& \vrule#& \strut #  
&\vrule#\tabskip=0pt\cr
\noalign{\centerline{Table \ttnic} \tableskip}
\noalign{\centerline
{\it Critical exponents: direct fit of $d=3$ series (error from 
${\tilde g}^*$ is not reported).}
\tableskip}
\fileth
height2.0pt
& \omit&& \omit&& \omit&& \omit&&\omit&\cr
&$ N,{\rm Ref.}\hfill$&
&$ \hfill \tilde{g}^* \hfill$&
&$ \hfill \gamma \hfill$&
&$ \hfill \nu \hfill$&
&$ \hfill \eta \hfill$&
\cr
height2.0pt& 
\omit&&\omit&&\omit&&  \omit&& \omit&\cr \fileth
height2.0pt& \omit&&\omit&&\omit&& \omit&& \omit&\cr 
&$\hfill 0, \rBGNMU \hfill$&&$ \hfill 1.39 \hfill$& &$ \hfill 1.1569\pm0.0004\hfill$&&$ \hfill
0.5872\pm 0.0004
\hfill$&&$ \hfill 0.0297\pm 0.0009 \hfill$&\cr 
&$\hfill 1, \rBGNMU \hfill$&&$ \hfill 1.40 \hfill$& &$ \hfill 1.2378\pm0.0006\hfill$&&$ \hfill
0.6301\pm 0.0005
\hfill$&&$ \hfill 0.0355\pm 0.0009 \hfill$&\cr 
&$\hfill 2, \rBGNMU \hfill$&&$ \hfill 1.40 \hfill$& &$ \hfill 1.3178\pm0.0010\hfill$&&$ \hfill
0.6715\pm 0.0007
\hfill$&&$ \hfill 0.0377\pm 0.0006 \hfill$&\cr
&$\hfill 3, \rBGNMU \hfill$&&$ \hfill 1.39 \hfill$& &$ \hfill 1.3926\pm0.0013\hfill$&&$ \hfill
0.7096\pm 0.0008
\hfill$&&$ \hfill 0.0374\pm 0.0004 \hfill$&\cr  
height2.0pt&\omit&&\omit&& \omit&& \omit&& \omit&\cr
\fileth}}$$
\endinsert

\midinsert
$$ \vbox{\elevenpoint\offinterlineskip\tabskip=0pt\halign to \hsize
{& \vrule#\tabskip=0em plus1em & \strut\ # \ 
& \vrule#& \strut #
& \vrule#& \strut #  
& \vrule#& \strut #  
& \vrule#& \strut #  
&\vrule#\tabskip=0pt\cr
\noalign{\centerline{Table \ttsok} \tableskip}
\noalign{\centerline
{\it Critical exponents: results of Pad\'e Borel summation for $N=4$,
\cite{\rAnSo}.}
\tableskip}
\fileth
height2.0pt
& \omit&& \omit&& \omit&& \omit&&\omit&\cr
&$ N, {\rm Ref.}\hfill$&
&$ \hfill \tilde{g}^* \hfill$&
&$ \hfill \gamma \hfill$&
&$ \hfill \nu \hfill$&
&$ \hfill \eta \hfill$&
\cr
height2.0pt& 
\omit&&\omit&&\omit&&  \omit&& \omit&\cr \fileth
height2.0pt& \omit&&\omit&&\omit&& \omit&& \omit&\cr 
&$\hfill 4, \rAnSo \hfill$&&$ \hfill 1.369 \hfill$& &$ \hfill 1.449
\hfill$&&$ \hfill
0.738
\hfill$&&$ \hfill 0.036 \hfill$&\cr
height2.0pt&\omit&&\omit&& \omit&& \omit&& \omit&\cr
\fileth}}$$
\endinsert


\midinsert
$$ \vbox{\elevenpoint\offinterlineskip\tabskip=0pt\halign to \hsize
{& \vrule#\tabskip=0em plus1em & \strut\ # \ 
& \vrule#& \strut #
& \vrule#& \strut #  
& \vrule#& \strut #  
& \vrule#& \strut #  
&\vrule#\tabskip=0pt\cr
\noalign{\centerline{Table \ttepsiold}\tableskip}
\noalign{\centerline{\it Estimates of critical exponents: 
previous  $\varepsilon$ expansion results (LG--ZJ).}
\tableskip} 
\fileth
height2.0pt& \omit&& \omit&& \omit&&\omit&& \omit&\cr
&$ \hfill N \hfill$&&$ \hfill 0
\hfill$&&$ \hfill 1 \hfill$&&$ \hfill 2
\hfill$&&$ \hfill 3 \hfill$&\cr
height2.0pt& \omit&& \omit&& \omit&& \omit&& \omit&\cr
\fileth
height2.0pt& \omit&& \omit&& \omit&& \omit&& \omit&\cr
&$ \hfill \gamma \hfill$&&$ \hfill 1.157
\pm0.003 \hfill$&&$ \hfill 1.2390\pm 0.0025
\hfill$&&$ \hfill 1.315\pm0.007  \hfill$&&$
\hfill 1.390\pm0.010 \hfill$&\cr
height2.0pt& \omit&& \omit&& \omit&& \omit&& \omit&\cr
&$ \hfill \nu \hfill$&&$ \hfill 0.5880\pm0.0015
 \hfill$&&$ \hfill 0.6310\pm0.0015
\hfill$&&$ \hfill 0.671\pm0.005  \hfill$&&$
\hfill 0.710\pm0.007 \hfill$&\cr
height2.0pt& \omit&& \omit&& \omit&& \omit&&\omit&\cr
&$ \hfill \eta \hfill$&&$ \hfill 0.0320\pm0.0025
 \hfill$&&$ \hfill 0.0375\pm0.0025
\hfill$&&$ \hfill 0.040\pm0.003  \hfill$&&$
\hfill 0.040\pm0.003 \hfill$&\cr
height2.0pt& \omit&& \omit&& \omit&& \omit&& \omit&\cr
&$ \hfill \beta \hfill$&&$ \hfill 0.3035\pm0.0020
 \hfill$&&$ \hfill 0.3270\pm0.0015
\hfill$&&$ \hfill 0.3485\pm0.0035  \hfill$&&$
\hfill 0.368\pm0.004 \hfill$&\cr
height2.0pt& \omit&& \omit&& \omit&& \omit&& \omit&\cr
&$ \hfill \omega \hfill$&&$ \hfill
0.82\pm0.04  \hfill$&&$ \hfill 0.81\pm0.04
\hfill$&&$ \hfill 0.80\pm 0.04 \hfill$&&$
\hfill 0.79 \pm0.04 \hfill$&\cr
height2.0pt& \omit&& \omit&& \omit&& \omit&& \omit&\cr
\fileth}}$$
\endinsert

\nref\rPelVic{A. Pellisetto and E. Vicari, preprint IFUP-TH 52/97,
cond-mat/9711078, data are taken from revised version submitted to Nucl. Phys. B.
} 
\midinsert
$$ \vbox{\elevenpoint\offinterlineskip\tabskip=0pt\halign to \hsize
{& \vrule#\tabskip=0em plus1em & \strut\ # \ 
& \vrule#& \strut #
& \vrule#& \strut #  
& \vrule#& \strut #  
& \vrule#& \strut #  
&\vrule#\tabskip=0pt\cr
\noalign{ \centerline{Table \ttpelvic}\tableskip}
\noalign{\centerline{\it Estimates of critical exponents from
$\varepsilon$-expansion \rPelVic.} \tableskip} 
\fileth
height2.0pt& \omit&& \omit&& \omit&&\omit&& \omit&\cr
&$ \hfill N \hfill$&&$ \hfill 0
\hfill$&&$ \hfill 1 \hfill$&&$ \hfill 2
\hfill$&&$ \hfill 3 \hfill$&\cr
height2.0pt& \omit&& \omit&& \omit&& \omit&& \omit&\cr
\fileth
height2.0pt& \omit&& \omit&& \omit&& \omit&& \omit&\cr
&$ \hfill \g^* (O(\varepsilon^4))\hfill$&
&$ \hfill1.390\pm0.017  \hfill$&
&$ \hfill 1.397\pm0.008\hfill$&
&$ \hfill 1.413\pm 0.013 \hfill$&
&$\hfill 1.387 \pm0.007 \hfill$&\cr
height2.0pt& \omit&& \omit&& \omit&& \omit&& \omit&\cr
&$ \hfill \gamma \hfill$&
&$ \hfill 1.1559\pm0.0010 \hfill$&
&$ \hfill 1.240\pm 0.005\hfill$&
&$ \hfill 1.304\pm0.007  \hfill$&
&$\hfill 1.372\pm0.006 \hfill$&
\cr
height2.0pt& \omit&& \omit&& \omit&& \omit&& \omit&\cr
&$ \hfill \nu \hfill$&
&$ \hfill 0.5882\pm0.0011\hfill$&
&$ \hfill 0.631\pm0.003\hfill$&
&$ \hfill 0.664\pm0.003  \hfill$&
&$\hfill 0.699\pm0.004 \hfill$&
\cr
height2.0pt& \omit&& \omit&& \omit&& \omit&& \omit&\cr
\fileth }}$$
\endinsert

\midinsert
$$ \vbox{\elevenpoint\offinterlineskip\tabskip=0pt\halign to \hsize
{& \vrule#\tabskip=0em plus1em & \strut\ # \ 
& \vrule#& \strut #  
& \vrule#& \strut #  
& \vrule#& \strut #  
& \vrule#& \strut #  
&\vrule#\tabskip=0pt\cr
\noalign{\centerline{Table \ttht} \tableskip}
\noalign{\centerline{\it Critical exponents for Ising-like systems: HT
series.} 
\tableskip} \fileth
height2.0pt& \omit&& \omit&& \omit&&\omit&& \omit&\cr
&$ N, {\rm Ref.}\hfill$&&$ \hfill \gamma \hfill$&&$ \hfill \nu \hfill$&&$
\hfill \alpha 
\hfill$&&$ \hfill \theta=\omega \nu \hfill $&\cr 
height2.0pt& 
\omit&&\omit&&\omit&& \omit&&\omit&\cr \fileth
height2.0pt& \omit&&\omit&&\omit&& \omit&& \omit&\cr 
&\hfill 0,\rBUCOM \hfill& &$ \hfill 1.1595\pm0.0012 \hfill$&&$ \hfill
0.588\pm0.001  
\hfill$&&$ \hfill \hfill$&&$ \hfill \hfill$&\cr
&\hfill 0,\rMacDo \hfill& &$ \hfill 1.16193\pm0.0001 \hfill$&&$ \hfill
0.588\pm0.001  
\hfill$&&$ \hfill \hfill$&&$ \hfill \hfill$&\cr
&\hfill 1,\rNickel \hfill& &$ \hfill 1.239\pm0.002 \hfill$&&$ \hfill
0.631\pm0.003  
\hfill$&&$ \hfill \hfill$&&$ \hfill \hfill$&\cr 
height2.0pt& \omit&&\omit&& \omit&& \omit&& \omit&\cr 
&\hfill 1,\rZJhts \hfill& &$ \hfill 1.2385\pm0.0025 \hfill$&&$
\hfill 0.6305\pm0.0015  
\hfill$&&$ \hfill \hfill$&&$
\hfill 0.57\pm0.07 \hfill$&\cr 
height2.0pt& \omit&&\omit&& \omit&& \omit&& \omit&\cr 
&\hfill 1,\rAdler \hfill& &$ \hfill 1.239\pm0.003 \hfill$&&$ \hfill
0.631\pm0.004  
\hfill$&&$ \hfill  \hfill$&&$ \hfill  \hfill$&\cr 
height2.0pt& \omit&&\omit&& \omit&& \omit&& \omit&\cr 
&\hfill 1,\rFiCh \hfill& &$ \hfill 1.2395\pm0.0004 \hfill$&&$ \hfill
0.632\pm0.001  
\hfill$&&$ \hfill 0.105\pm0.007 \hfill$&&$
\hfill 0.54 \pm0.05 \hfill$&\cr 
height2.0pt& \omit&&\omit&& \omit&& \omit&& \omit&\cr 
&\hfill 1,\rGutt  \hfill& &$ \hfill 1.239\pm0.003 \hfill$&&$ \hfill
0.632\pm0.003  
\hfill$&&$ \hfill 0.101\pm.004 \hfill$&&$ \hfill  \hfill$&\cr 
height2.0pt& \omit&&\omit&& \omit&& \omit&& \omit&\cr 
&\hfill 1,\rNiRe\hfill& &$ \hfill 1.237\pm0.002 \hfill$&&$ \hfill
0.630\pm0.0015  \hfill$&&$ \hfill  \hfill$&&$
\hfill 0.52\pm0.03 \hfill$&\cr
 height2.0pt&\omit&&\omit&& \omit&&
\omit&& \omit&\cr
& \hfill 1,\rBCGS \hfill& &$ \hfill 
\hfill$&&$ \hfill  \hfill$&&$ \hfill 0.104\pm0.004 \hfill$&&$
\hfill  \hfill$&\cr
&\hfill 1,\rBUCOM \hfill& &$ \hfill 1.2385\pm0.0005 \hfill$&&$ \hfill
0.6310 \pm0.0005  
\hfill$&&$ \hfill \hfill$&&$ \hfill \hfill$&\cr 
&\hfill 1,\rKosuzu \hfill& &$ \hfill 1.237\pm0.004 \hfill$&&$ \hfill
  \hfill$&&$ \hfill .108\pm.005\hfill$&&$ \hfill \hfill$&\cr
&\hfill 2,\rBUCOM \hfill& &$ \hfill 1.323\pm0.003 \hfill$&&$ \hfill
0.674\pm0.003  
\hfill$&&$ \hfill \hfill$&&$ \hfill \hfill$&\cr
&\hfill 2,\FeMoW \hfill& &$ \hfill 1.323\pm0.015 \hfill$&&$ \hfill
0.670\pm0.007  
\hfill$&&$ \hfill \hfill$&&$ \hfill \hfill$&\cr
&\hfill 3,\rBUCOM \hfill& &$ \hfill 1.402\pm0.003 \hfill$&&$ \hfill
0.714\pm0.002  
\hfill$&&$ \hfill \hfill$&&$ \hfill \hfill$&\cr
&\hfill 3,\RiFiAl\hfill& &$ \hfill 1.40\pm0.03 \hfill$&&$ \hfill
0.72\pm0.01  
\hfill$&&$ \hfill \hfill$&&$ \hfill \hfill$&\cr
&\hfill 4,\rBUCOM \hfill& &$ \hfill 1.474\pm0.004 \hfill$&&$ \hfill
0.750\pm0.003  
\hfill$&&$ \hfill \hfill$&&$ \hfill \hfill$&\cr
height2.0pt&\omit&& \omit&& \omit&& \omit&& \omit&\cr
\fileth}}$$
\endinsert
\nref\rBeloNick{P. Belohorec and B.G. Nickel, {\it Accurate universal and
two-parameter model results from a Monte-Carlo renormalization group study},
Guelph Univ. preprint 27/09/97.}
\midinsert
$$ \vbox{\elevenpoint\offinterlineskip\tabskip=0pt\halign to \hsize
{& \vrule#\tabskip=0em plus1em & \strut\ # \ 
& \vrule#& \strut #
& \vrule#& \strut #  
& \vrule#& \strut #  
& \vrule#& \strut #  
& \vrule#& \strut #  
&\vrule#\tabskip=0pt\cr
\noalign{\centerline{Table \ttmc} \tableskip}
\noalign{\centerline
{\it Critical exponents: MC}
\tableskip} \fileth
height2.0pt& \omit&& \omit&& \omit&& \omit&&\omit&& \omit&\cr
&$ N, {\rm Ref.}\hfill$&&$ \hfill \gamma \hfill$&&$ \hfill \nu \hfill$&&$
\hfill \beta 
\hfill$&&$ \hfill \eta\hfill$&&$ \hfill \theta=\omega \nu \hfill $&\cr
height2.0pt& \omit&&\omit&&\omit&&  \omit&& \omit&&\omit&\cr \fileth
height2.0pt& \omit&&\omit&&\omit&& \omit&& \omit&& \omit&\cr 
&$\hfill 0, \cite{\rcacape,\rsokal} \hfill$& &$ \hfill 1.1575\pm 0.0006 \hfill$&&$ \hfill 0.5877\pm 0.0006  \hfill$&&$ \hfill \hfill$&&$ \hfill  \hfill$&&$\hfill \hfill$&
\cr 
&$\hfill 0, \rBeloNick \hfill$& &$ \hfill  \hfill$&&$ \hfill 0.58758\pm
0.00007  \hfill$&&$ \hfill \hfill$&&$ \hfill
\hfill$&&$\hfill0.515{+.017\atop-.007} \hfill$&  
\cr 
&$\hfill 1, \cite{\rBloet} \hfill$& &$ \hfill \hfill$&&$ \hfill 0.631\pm 0.001  \hfill$&&$ \hfill 0.3269\pm0.0006\hfill$&&$ \hfill .038\pm .002 \hfill$&&$\hfill  \hfill$&
\cr 
&$\hfill 1, \cite{\rFeLa} \hfill$& &$ \hfill \hfill$&&$ \hfill 0.6289
\pm 0.0008  \hfill$&&$ \hfill \hfill$&&$ \hfill  \hfill$&&$\hfill  \hfill$&
\cr 
&$\hfill 1, \cite{\rRaGu} \hfill$& &$ \hfill \hfill$&&$ \hfill 0.625\pm 0.001  \hfill$&&$ \hfill \hfill$&&$ \hfill  .0025\pm.006\hfill$&&$\hfill .44 \hfill$&
\cr 
&$\hfill 1, \cite{\rBFMM} \hfill$& &$ \hfill  \hfill$&&$ \hfill 0.6294\pm.0009 \hfill$&&$ \hfill \hfill$&&$ \hfill .0374 \pm.0014 \hfill$&&$\hfill  .55 \pm .06\hfill$&
\cr
&$\hfill 2, \cite{\rGHM} \hfill$& &$ \hfill 1.324\pm.001 \hfill$&&$ \hfill 0.664\pm.006  \hfill$&&$ \hfill \hfill$&&$ \hfill  \hfill$&&$\hfill   \hfill$&
\cr 
&$\hfill 2, \cite{\rJanke} \hfill$& &$ \hfill 1.323\pm.002 \hfill$&&$ \hfill 0.670\pm.002 \hfill$&&$ \hfill \hfill$&&$ \hfill  \hfill$&&$\hfill  \hfill$&
\cr 
&$\hfill 2, \cite{\rBFMM} \hfill$& &$ \hfill 1.316\pm.003 \hfill$&&$ \hfill 0.6721\pm.0013 \hfill$&&$ \hfill \hfill$&&$ \hfill .042 \pm.002 \hfill$&&$\hfill  .54 \pm .08\hfill$&
\cr 
&$\hfill 3, \cite{\rJAHO} \hfill$& &$ \hfill 1.3896\pm.0070 \hfill$&&$ \hfill 0.7036\pm.0023 \hfill$&&$ \hfill.362\pm .004 \hfill$&&$ \hfill .0027\pm0.002 \hfill$&&$\hfill  \hfill$&
\cr 
&$\hfill 3, \cite{\rBFMM} \hfill$& &$ \hfill 1.396\pm.003 \hfill$&&$ \hfill 0.7128\pm.0014 \hfill$&&$ \hfill \hfill$&&$ \hfill  .041\pm .002\hfill$&&$\hfill  .51 \pm.11\hfill$&
\cr
&$\hfill 4, \cite{\rBFMM} \hfill$& &$ \hfill 1.476\pm.002 \hfill$&&$ \hfill 0.7525\pm.0010 \hfill$&&$ \hfill \hfill$&&$ \hfill .038 \pm.001 \hfill$&&$\hfill  \hfill$&
\cr  
&$\hfill 4, \cite{\rKAKA} \hfill$& &$ \hfill 1.477\pm.018 \hfill$&&$ \hfill 
0.748\pm.009 \hfill$&&$ \hfill .3836\pm.0046\hfill$&&$ \hfill  \hfill$&&$\hfill  \hfill$&
\cr   
height2.0pt&\omit&&\omit&& \omit&& \omit&& \omit&& \omit&\cr
\fileth}}$$
\endinsert


\midinsert
$$ \vbox{\elevenpoint\offinterlineskip\tabskip=0pt\halign to \hsize
{& \vrule#\tabskip=0em plus1em & \strut\ # \ 
& \vrule#& \strut #
& \vrule#& \strut #  
& \vrule#& \strut #  
& \vrule#& \strut #  
& \vrule#& \strut #  
&\vrule#\tabskip=0pt\cr
\noalign{\centerline{Table \tterg} \tableskip}
\noalign{\centerline
{\it Critical exponents: ``Exact Renormalization Group" estimates.}
\tableskip} \fileth
height2.0pt& \omit&& \omit&& \omit&& \omit&&\omit&& \omit&\cr
&$ N, {\rm Ref.}\hfill$&&$ \hfill \tilde{g}^* \hfill$&&$ \hfill \gamma
\hfill$&&$ \hfill \nu \hfill$& 
&$ \hfill \eta \hfill$&&$ \hfill \theta=\omega \nu \hfill $&\cr
height2.0pt& 
\omit&&\omit&&\omit&&  \omit&& \omit&&\omit&\cr \fileth
height2.0pt& \omit&&\omit&&\omit&& \omit&& \omit&& \omit&\cr 
&$\hfill 1, \rwetterich \hfill$&&$ \hfill 1.726\hfill$& &$ \hfill 
1.247\hfill$&&$ \hfill 0.638
\hfill$&&$ \hfill 0.045\hfill$&&$
\hfill  \hfill$&\cr 
&$\hfill 2, \rwetterich \hfill$&&$ \hfill 1.675\hfill$& &$ \hfill
1.371\hfill$&&$ \hfill 0.700\hfill$&&$ \hfill 0.042 \hfill$&&$
\hfill  \hfill$&\cr 
&$\hfill 3, \rwetterich \hfill$&&$ \hfill 1.619\hfill$& &$ \hfill
1.474\hfill$&&$ \hfill 0.752 \hfill$&&$ \hfill 0.038 \hfill$&&$
\hfill  \hfill$&\cr 
&$\hfill 4, \rwetterich \hfill$&&$ \hfill 1.566\hfill$& &$ \hfill
1.556\hfill$&&$ \hfill 0.791 \hfill$&&$ \hfill 0.034 \hfill$&&$
\hfill  \hfill$&\cr 
&$\hfill 1, \rmorris \hfill$&&$ \hfill\hfill$& &$ \hfill 
\hfill$&&$ \hfill 
0.618\pm 0.014
\hfill$&&$ \hfill .054\hfill$&&$
\hfill  0.56\pm 0.07\hfill$&\cr 
&$\hfill 1, \rWRolf \hfill$&&$ \hfill \hfill$& &$ \hfill \hfill$&&$ \hfill
0.6262\pm 0.0013
\hfill$&&$ \hfill \hfill$&&$
\hfill  \hfill$&\cr 
&$\hfill 1, \rTravesset \hfill$&
&$ \hfill \hfill$& 
&$ \hfill \hfill$&
&$ \hfill 0.625\pm 0.007\hfill$&
&$ \hfill 0.030\pm 0.005 \hfill$&&$
\hfill 0.48\pm 0.04 \hfill$&\cr 
height2.0pt&\omit&&\omit&& \omit&& \omit&& \omit&& \omit&\cr
\fileth}}$$
\endinsert

\nref\rTravesset{A. Travesset, hep-lat 9709094, {\it Nucl. Phys.} B
(Proc. Suppl.) 63A-C (1998) 640.}

\midinsert
$$ \vbox{\elevenpoint\offinterlineskip\tabskip=0pt\halign to \hsize
{& \vrule#\tabskip=0em plus1em & \strut\ # \ 
& \vrule#& \strut #
& \vrule#& \strut #  
& \vrule#& \strut #  
& \vrule#& \strut #  
& \vrule#& \strut #  
&\vrule#\tabskip=0pt\cr
\noalign{\centerline{Table \ttEXP} \tableskip}
\noalign{\centerline
{\it Critical exponents: selected recent experiments}
\tableskip} \fileth
height2.0pt& \omit&& \omit&& \omit&& \omit&&\omit&& \omit&\cr
&$ N, {\rm Ref.}\hfill$&&$ \hfill \gamma \hfill$&&$ \hfill \nu \hfill$&&$
\hfill \beta 
\hfill$&&$ \hfill \alpha\hfill$&&$ \hfill \theta=\omega \nu \hfill $&
\cr
height2.0pt& \omit&&\omit&&\omit&&  \omit&& \omit&&\omit&\cr \fileth
height2.0pt& \omit&&\omit&&\omit&& \omit&& \omit&& \omit&\cr 
&$\hfill 0 ,\cite{\rCot} \hfill$& &$ \hfill \hfill$&&$ \hfill 0.586\pm 0.004  \hfill$&&$ \hfill  \hfill$&&$ \hfill \hfill$&&$\hfill \hfill$&
\cr
&$\hfill 1,\cite{\rFlede} \hfill$& &$ \hfill \hfill$&&$ \hfill   \hfill$&&$ \hfill  \hfill$&&$ \hfill 0.107\pm0.006 \hfill$&&$\hfill \hfill$&
\cr
&$\hfill 1,\cite{\rRaKiJa,\rBelan,\rHenkel} \hfill$& &$ \hfill 1.25\pm0.01 \hfill$&&$ \hfill 0.64\pm 0.01  \hfill$&&$ \hfill  \hfill$&&$ \hfill  0.109\pm0.006\hfill$&&$\hfill 0.57 \pm .09\hfill$&
\cr
&$\hfill 1,\cite{\rPestakcha} \hfill$& &$ \hfill 1.233\pm0.010 \hfill$&& &&$ \hfill 0.327\pm0.002 \hfill$&& &&$\hfill 0.51 \pm 0.03\hfill$&
\cr

&$\hfill 2, \cite{\rLipa,\rLIpabis} \hfill$& 
&$ \hfill \hfill$&&$ \hfill 0.6708 \pm 0.0004 \hfill$&&$ \hfill   \hfill$&&$ 
\hfill -.01285\pm0.00038  \hfill$&&$\hfill
 \hfill$&
\cr
&$\hfill 2, \cite{\rgomuah} 
\hfill$& 
&$ \hfill \hfill$&&$ \hfill 0.6705 \pm 0.0006 \hfill$&&$ \hfill   \hfill$&&$ 
\hfill  \hfill$&&$\hfill
 \hfill$&
\cr
&$\hfill 3, \cite{\rHenkel} 
\hfill$& 
&$ \hfill \hfill$&&$ \hfill  \hfill$&&$ \hfill   \hfill$&&$ 
\hfill \hfill$&&$\hfill
  0.61 \pm0.06\hfill$&
\cr
height2.0pt&\omit&&\omit&& \omit&& \omit&& \omit&& \omit&\cr
\fileth}}$$
\endinsert

\nref\rFlede{
A.C. Flewelling, R.J. Defonseka, N. Khaleeli, J. Partee and D.T.
Jacobs, {\it J. Chem. Phys.} 104 (1996) 8048.
}

\nref\rRaKiJa{
C.A. Ramos, A.R. King and V. Jaccarino
{\it Phys. Rev. B}10 (1989) 7124}

\nref\rBelan{D.P. Belanger and H. Yoshizawa, {\it Phys. Rev.B} 35(1987) 4823.}

\nref\rPestakcha{M.W. Pestak and H.W. Chan {\it Phys. Rev B} 30 (1984) 274.} 

\nref\rLIpabis{D.R. Swanson, T.C.P.Chui and J.A. Lipa, 
{\it Phys. Rev B}46 (1992) 9043.
D. Marek, J.A. Lipa and D. Philips, {\it Phys. Rev B}38 (1988) 4465.  }
\nref\rgomuah{L.S. Goldner, N. Mulders and G. Ahlers, J. Low Temp.Phys. 93
(1992) 131.} 

\nref\rHenkel{M. Henkel, S. Andrieu, P. Bauer and M. Piecuch,
{\it Phys. Rev. Lett.} 80 (1998) 4783.}


\par
For what concerns experimental determinations of critical exponents
a few significant results are displayed in Table \ttEXP.

\medskip
{\it 3D series.}
In general the new estimates displayed in Table \ttd3 are more accurate than
the previous LG-ZJ results. They are compatible within errors with the
previous analysis. A closer inspection shows however some significant changes
which require discussion. \par
The main effect comes from the new (and smaller)  values of the fixed
point coupling constant for $N<3$. The changes are a direct
consequence of the new techniques we have introduced. In the old
calculation LG--ZJ had noticed two puzzling features: the optimal values of
the parameter $b$ were
somewhat large, compared to what large order behaviour did suggest. Moreover
the three shifts $0,1,2$ gave strongly oscillating results.\par
Introduction of the new parameter $r$ (see Eq.~{\eerre}) has shown that the
old apparent convergence 
corresponded to an unstable region of parameters. By varying $r$
we find a region where these problems are solved to a large extent: the
results are less sensitive, various shifts agree, and all parameters have
more reasonable values. Figure 1 exemplifies this situation for the 
$N=0$ case.
\par
\midinsert
\vskip-6cm
\epsfysize=18cm
\epsfxsize=12cm
\centerline{\epsfbox{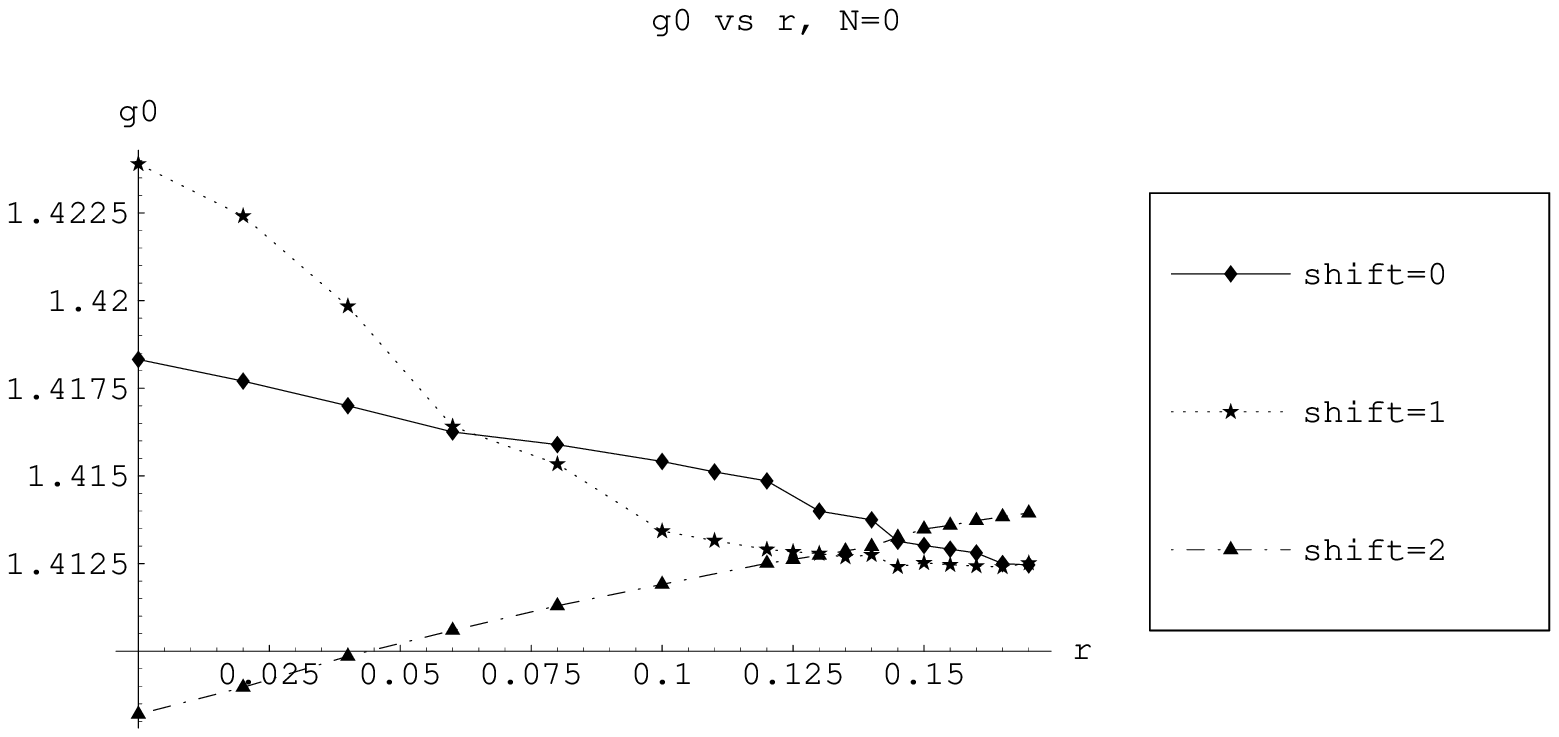}}
\vskip-6cm
\figure{3.mm}{Values of $g^*$, $N=0$, as a function of the parameter $r$
for shifts 0,1,2}
\endinsert
Another example exhibited a similar instability at $r=0$: 
$\gamma$, $N=0$ $d=3$.
This (as well as the decrease induced from that of $g^*$) explains
the new different value we obtained here.\par
Finally $N=3$ values show a consistent effect: the three exponents
$\gamma,\nu,\beta$ increase. This simply suggests that $N=3$
errors had been underestimated in LG--ZJ analysis.\par
Conversely the values of $\nu$, $N=0,1,2$ are quite stable, even though the
corresponding values of $g^*$ have changed. The reasons are a very good
apparent convergence of the pseudo-epsilon expansion (on which previous
analysis partially relied) at previous order.\par 
The apparent errors have generally been reduced, as should be expected,
except for $N=3$ (see the comment above). 
The improvement is specially significant for the exponent $\eta$ that was
poorly determined before. With a few exceptions the general trend for a given
exponent is the increase of the direct summation error with 
$N$. This effect has a simple explanation: our summation method relies on 
the large order behaviour analysis, and the asymptotic regime sets in later
when $N$ increases \ref\rPABRE{E. Br\'ezin and G. Parisi, {\it J. Stat.
Phys.} 19 (1978) 269.}. Perhaps a clever use of the knowlege coming from the
large $N$ expansion could improve the situation.\par
Note finally that the term added to two of the three RG series has allowed not
only to decrease apparent errors but also to estimate them more reliably.\par
For several exponents (such as e.g.~$\gamma(N=1)$) errors are now 
dominated by errors induced by the determination of $g^*$. 
To further improve the situation it will be necessary to also add a new term
to the RG $\beta$-function. 
\medskip
{\it (Free) $\varepsilon$-expansion  and 3D series.} Comparison with the
previous LG--ZJ $\varepsilon$ estimates shows no striking effect, small
deviations being due to the use of corrected series.\par  
For the exponent $\nu$ the consistency between 3D and $\varepsilon$ results
remains very good at all $N$. The situation has markedly improved for the
exponent $\gamma$, $N=0$: there is still a systematic discrepancy in
the central values for the exponents (about $0.002$), but the difference is
reduced by more than a factor two, which is quite encouraging.
A similar comment applies to the central values of the exponent $\eta$
$N=0,\cdots, 3$, where the discrepancy is also reduced by a factor two.  
For what concerns the $N=4$ prediction, the agreement with the corresponding
$d=3$ results is quite satisfactory but apparent errors are large.\par
One point should however be stressed: since the series are shorter it is more
difficult to assess apparent errors and the errors we quote are thus less
reliable than for the $d=3$ series. 
\medskip
{\it Free and bc $\varepsilon$-expansion.}
For the $\varepsilon$-expansion we report a second  set of values,
obtained by imposing the exact $d=2$ values,
referred as bc (i.e.~with boundary conditions) in Table \tteps~to distinguish
them from the unconstrainted values denoted by free.
Some remarks are in order about the bc approach.
First we do not know the analytic structure of exponents 
near $d=2$. The only piece of
evidence comes from using the $\varepsilon$-expansion for $d=2$.
For $N=0$ the agreement with exact results is quite good \rLGZJiii,
suggesting that $d=2$ is a regular point. For $N=1$ the agreement is less
striking, leaving room for some complex behaviour. For values $N>2$
there are indications (IR-renormalons of the $d=2$ non-linear $\sigma$-model)
that $d=2$ corresponds to an essential  singularity (probably
non-Borel summable). It is then easy to construct examples where 
fitting leads to worse results.\par
Second  in the case $N\ge 2$ the new series have a more complicated structure
which makes summation (which is a form of extrapolating
series to higher orders) and even more error estimation difficult. 
\par
Although the two analyses give compatible results,
it happens that  bc has the general
tendency  to give  values of $\gamma, \nu$  for $N\ge2$ 
larger than free ones and values of $\eta$  for $N>2$ 
smaller than free ones. This point is  understood
by remembering that at $d=2$ the corresponding $\gamma, \nu, 1/\eta$
are infinite and thus it is reasonable that imposing this behaviour at $d=2$
tends to increase the value of the exponent at $d=3$.  
\par 
Finally it is also remarkable that for $N=0$ the bc errors are smaller than
the free ones. 

\medskip
{\it Other methods.} The general agreement between the HT series, the MC
results, and the new $d=3$ determinations is in general improved.
This in particular applies to the SAW where recent long simulations provide
very accurate estimates. 

\midinsert
\epsfysize=7cm
\epsfxsize=12cm
\centerline{\epsfbox{nu2.epsf}}
\vskip-1cm
\figure{3.mm}{Comparison between various estimates of the exponent $\nu$,
N=2:
$He\ li$ from \cite{\rLipa,\rLIpabis},
$He\ go$ from \cite{\rgomuah}, 
$d3\ gz$ from present work ($d=3$), 
$d3\ lz$ from  LG--ZJ ($d=3$), 
$ep\ gz$ from present work ($\varepsilon$), 
$ep\ lz$ from  LG{--}ZJ ($\varepsilon$),
$ep\ pv$ from  \cite{\rPelVic} ($\varepsilon$), 
$mc\ j$ from \cite{\rJanke},
$mc\ b$ from \cite{\rBFMM},
$ht\ bc$ from  \cite{\rBUCOM}.
}  
\endinsert
\medskip
{\it Experiments.} The improved agreement of present results 
with the recent measures on superfluid Helium systems, $N=2$, is
remarkable and is displayed graphically in Figure 2.
In spite of our efforts the best experimental value is still much more 
accurate than the theoretical estimate. In all other cases the agreement with
experiments is good as seen from Table \ttEXP.

\medskip
{\bf Acknowledgments.} The authors are indebted to 
 Bernie G. Nickel for sending them copy of his preprint and to
H. Kleinert for providing a file with $\varepsilon $ expansion results.
One of the author (R.~G.) wants also to thank 
 the INFN group of Genova for its kind
hospitality. The work of R.~G.~is supported by an EC TMR grant, 
contract N$^o$ ERB-FMBI-CT-95.0130. 
%
\listrefs
\vfill
\eject
\def\appendixname{Series $d=3$}
\appendix{}

\section{\appendixname}

We report here the seven loop series for the $O(N)$ symmetric $(\phi^2)^2$ 
theory, $N=0,1,2,3$, computed by Murray and Nickel \cite{\rBGNMU}.
The functions $\eta(\g)$ $\eta_2(\g)$ below are defined by
$$
\eta(\g)= m{d \log Z \over \d m } \qquad \eta_2(\g)= m{d \log Z_2 \over \d m }
$$
where the renormalization constant are defined by
$\phi_0=\sqrt{Z} \phi_r$ and $(\phi^2)_r={Z_2\over Z} (\phi_0)^2$
(subscript $0$,$r$ indicate respectively bare adn renormalized fields).
The critical exponents $\eta, \nu$ can be found by the identification
$\eta=\eta(\g^*)$ and $\nu=(2+\eta_2(\g^*)-\eta(\g^*))^{-1}$.
The symmetry number $N$ is reported in square brackets below.
\sslbl\ssapp

$$
\eqalignno{ 
\eta[0] &= {{{\g^2}}\over {108}} + 0.0007713750{\g^3} + 0.0015898706{\g^4}
 -0.0006606149{\g^5} \cr &+ 0.0014103421{\g^6} - 0.001901867{\g^7}\cr
 \cr
\eta[1] &= {{8{\g^2}}\over {729}} + 0.0009142223{\g^3} + 0.0017962229{\g^4} - 
    0.0006536980{\g^5}\cr & + 0.0013878101{\g^6} - 0.001697694{\g^7}\cr
 \cr
\eta[2] &= {{8{\g^2}}\over {675}} + 0.0009873600{\g^3} + 0.0018368107{\g^4} - 
    0.0005863264{\g^5}\cr & + 0.0012513930{\g^6} - 0.001395129{\g^7}\cr
 \cr
\eta[3] &= {{40{\g^2}}\over {3267}} + 0.0010200000{\g^3} + 0.0017919257{\g^4} - 
    0.0005040977{\g^5}\cr & + 0.0010883237{\g^6} - 0.001111499{\g^7}\cr
}
$$
$$
\eqalignno{ 
{ {\eta_2}}[0] &= 
   {{-\g}\over 4} + {{{\g^2}}\over {16}} - 0.0357672729{\g^3} + 
    0.0343748465{\g^4} - 0.0408958349{\g^5} \cr &+ 0.0597050472{\g^6} - 
    0.09928487{\g^7}\cr
 \cr
{ {\eta_2}}[1] &= 
   {{-\g}\over 3} + {{2{\g^2}}\over {27}} - 0.0443102531{\g^3} + 
    0.0395195688{\g^4} - 0.0444003474{\g^5} \cr &+ 0.0603634414{\g^6} - 
    0.09324948{\g^7}\cr
 \cr                                                                      
{ {\eta_2}}[2] &= 
   {{-2\g}\over 5} + {{2{\g^2}}\over {25}} - 0.0495134446{\g^3} + 
    0.0407881055{\g^4} - 0.0437619509{\g^5} \cr &+ 0.0555575703{\g^6} - 
    0.08041336{\g^7}\cr
 \cr
{ {\eta_2}}[3] &= 
   {{-5\g}\over {11}} + {{10{\g^2}}\over {121}} - 0.0525519564{\g^3} + 
    0.0399640005{\g^4} - 0.0413219917{\g^5} \cr &+ 0.0490929344{\g^6} - 
    0.06708630{\g^7}\cr
}
$$

\draftend
\end

%% file: sacmace.tex
\input hyperbasics
\catcode`\@=11
\def\unredoffs{\voffset=13mm \hoffset=6.5truemm} 
\def\redoffs{\voffset=-12.truemm\hoffset=-9truemm} 
\def\speclscape{}
%
\newbox\leftpage \newdimen\fullhsize \newdimen\hstitle \newdimen\hsbody
\newdimen\hdim
\hfuzz=1pt
\ifx\hyperdef\UNd@FiNeD\def\hyperdef#1#2#3#4{#4}\def\hyperref#1#2#3#4{#4}\fi
\def\newans{y }
\def\answb{y }
\ifx\answb\newans\message{(This uses normal fonts.)}%
%
\def\bigans{b }
\def\answ{b }
\ifx\answ\bigans\message{(Format simple colonne 12pts.}
\magnification=1200 \unredoffs\hsize=147truemm\vsize=219truemm
\hsbody=\hsize \hstitle=\hsize 
\else\message{(Format double colonne, 10pts.} \let\l@r=L
\magnification=1000 \vsize=182.5truemm
\redoffs%
\hstitle=122.5truemm\hsbody=122.5truemm\fullhsize=261.5truemm\hsize=\hsbody 
\output={
  \almostshipout{\leftline{\vbox{\makeheadline\pagebody\makefootline}}}
\advancepageno%
}
\def\almostshipout#1{\if L\l@r \count1=1 \message{[\the\count0.\the\count1]}
      \global\setbox\leftpage=#1 \global\let\l@r=R
 \else \count1=2
  \shipout\vbox{\speclscape{\hsize\fullhsize}
      \hbox to\fullhsize{\box\leftpage\hfil#1}}  \global\let\l@r=L\fi}
\fi

\input lfont
\def\sla#1{\mkern-1.5mu\raise0.4pt\hbox{$\not$}\mkern1.2mu #1\mkern 0.7mu}
\def\Dbar{\mkern-1.5mu\raise0.4pt\hbox{$\not$}\mkern-.1mu {\rm D}\mkern.1mu}
\def\Abar{\mkern1.mu\raise0.4pt\hbox{$\not$}\mkern-1.3mu A\mkern.1mu}
\nopagenumbers
\headline={\ifnum\pageno=1\hfill\else\draftdate\hfil{\headrm\folio}%
\hfil\fi}	 
\else\message{(This uses pseudo 12pts fonts.}
\hoffset=8mm
\voffset=16mm
\input lfont12 

\def\sla#1{\mkern-1.5mu\raise0.5pt\hbox{$\not$}\mkern1.2mu #1\mkern 0.7mu}
\def\Dbar{\mkern-1.5mu\raise0.5pt\hbox{$\not$}\mkern-.1mu {\rm D}\mkern.1mu}
\def\Abar{\mkern1.mu\raise0.5pt\hbox{$\not$}\mkern-1.3mu A\mkern.1mu}
\fi
\def\fileth{\noalign{\hrule}}

\newcount\yearltd\yearltd=\year\advance\yearltd by -1900
\newif\ifdraftmode
\draftmodefalse
\def\draft{\draftmodetrue{\count255=\time\divide\count255 by 60
\xdef\hourmin{\number\count255} 
  \multiply\count255 by-60\advance\count255 by\time
  \xdef\hourmin{\hourmin:\ifnum\count255<10 0\fi\the\count255}}}
\def\draftdate{\ifdraftmode{\headrm\quad (le
\number\day/\number\month/\number\yearltd\ \ \hourmin)}\else{}\fi} 
\newif\iffrancmode
\francmodefalse
\def\e{\mathop{\rm e}\nolimits}

\def\d{{\rm d}}
\def\ud{{\textstyle{1\over 2}}}

\chardef\sigmat=27

\def\frac#1#2{{\textstyle{#1\over#2}}}

\def\today{\number\day/\number\month/\number\year}
\def\leaderfill{\leaders\hbox to 1em{\hss.\hss}\hfill}
\catcode`\@=11
\def\deqalignno#1{\displ@y\tabskip\centering \halign to
\displaywidth{\hfil$\displaystyle{##}$\tabskip0pt&$\displaystyle{{}##}$
\hfil\tabskip0pt &\quad
\hfil$\displaystyle{##}$\tabskip0pt&$\displaystyle{{}##}$ 
\hfil\tabskip\centering& \llap{$##$}\tabskip0pt \crcr #1 \crcr}}
\def\deqalign#1{\null\,\vcenter{\openup\jot\m@th\ialign{
\strut\hfil$\displaystyle{##}$&$\displaystyle{{}##}$\hfil
&&\quad\strut\hfil$\displaystyle{##}$&$\displaystyle{{}##}$
\hfil\crcr#1\crcr}}\,}
\newread\ch@ckfile
\def\cinput#1{\def\filen@me{#1}
\immediate\openin\ch@ckfile=\filen@me
\ifeof\ch@ckfile\closein\ch@ckfile\message{<< (\filen@me\ N'EXISTE PAS)
>>}\else%
\input\filen@me\closein \ch@ckfile\fi}
\immediate\openin\ch@ckfile=\jobname.def
\ifeof\ch@ckfile\closein\ch@ckfile\message{<< (\jobname.def N'EXISTE PAS)
>>}
\def\DefWarn#1{\ifx\UNd@FiNeD#1\else
\immediate\write16{*** WARNING: the label \string#1 is already defined%
***}\fi}%
\else%
\def\DefWarn#1{}
\input\jobname.def\closein \ch@ckfile\fi
\newcount\nosection
\newcount\nosubsection
\newcount\neqno
\newcount\notenumber
\newcount\nofigure
\newif\ifappmode
\def\table{\jobname.toc}
\def\equation{\jobname.equ}
\def\labeldefs{\jobname.eqd}
\newwrite\equa
\newwrite\tab 
\newwrite\eqdf

\newdimen\hulp
\def\maketitle#1{
\edef\oneliner##1{\centerline{##1}}
\edef\twoliner##1{\vbox{\parindent=0pt\leftskip=0pt plus 1fill\rightskip=0pt
plus 1fill 
                     \parfillskip=0pt\relax##1}} 
\setbox0=\vbox{#1}\hulp=0.5\hsize
                 \ifdim\wd0<\hulp\oneliner{#1}\else
                 \twoliner{#1}\fi}
\def\preprint#1{{\sacfont }\hfill{#1}\vskip 20mm}
\def\title#1\par{\gdef\titlename{#1}
\maketitle{\ssbx\uppercase\expandafter{\titlename}}
\vskip20truemm
\nosection=0
\neqno=0
\notenumber=0
\nofigure=0
\def\prefix{}
\appmodefalse
\mark{\the\nosection}
\message{#1}
\immediate\openout\equa=\equation
\immediate\openout\eqdf=\labeldefs
}
\def\abstract{\vskip8mm\iffrancmode\centerline{RESUME}\else%
\centerline{ABSTRACT}\fi \smallskip \begingroup\narrower
\elevenpoint\baselineskip10pt} 
\def\endabstract{\par\endgroup \bigskip}
\def\section#1\par{\vskip0pt plus.1\vsize\penalty-100\vskip0pt plus-.1
\vsize\bigskip\vskip\parskip
\ifnum\nosection=0\ifappmode\relax\else\writetoc
\fi\fi
\advance\nosection by 1\global\nosubsection=0\global\neqno=0
\vbox{\noindent\bf{\hyperdef\hypernoname{section}{\prefix\the\nosection}%
{\prefix\the\nosection}\ #1}}
\writetoca{{\string\hyperref{}{section}{\prefix\the\nosection}%
{\prefix\the\nosection}} {#1}}
\message{\the\nosection\ #1}
\mark{\the\nosection}\bigskip\noindent
}

\def\appendix#1#2\par{\bigbreak\nosection=0
\appmodetrue
\notenumber=0
\neqno=0
\def\prefix{A}
\mark{\the\nosection}
\message{APPENDICES}
{\leftline{APPENDICES} \hyperdef\hypernoname{appendix}{app}{ 
\leftline{\uppercase\expandafter{#1}}
\leftline{\uppercase\expandafter{#2}}}}
\bigskip\noindent\nonfrenchspacing
\writetoca{\string\hyperref{}{appendix}{app}{Appendices}.\ #1.\ #2}%
}
\def\subsection#1\par {\vskip0pt plus.05\vsize\penalty-100\vskip0pt
plus-.05\vsize\bigskip\vskip\parskip\advance\nosubsection by 1
\vbox{\noindent\it{\hyperdef\hypernoname{subsection}{\prefix\the\nosection.%
\the\nosubsection}{\prefix\the\nosection.\the\nosubsection\ #1}}}%
\smallskip\noindent 
\writetoca{{\string\hyperref{}{subsection}{\prefix\the\nosection.%
\the\nosubsection}{\prefix\the\nosection.\the\nosubsection}} {#1}}
\message{\the\nosection.\the\nosubsection\ #1}
} 
\def\note #1{\advance\notenumber by 1
\footnote{$^{\the\notenumber}$}{\sevenrm #1}} 

\parindent=1em 
\newinsert\margin
\dimen\margin=\maxdimen
\count\margin=0 \skip\margin=0pt
\def\sslbl#1{\DefWarn#1%
\ifdraftmode{\hfill\escapechar-1{\rlap{\hskip-1mm%
\sevenrm\string#1}}}\fi%
\ifnum\nosection=0\xdef#1{}%
\edef\ewrite{\write\eqdf{\string\def\string#1{}}
\write\eqdf{}}\ewrite%
\edef\ewrite{\write\equa{{\string#1}}%
\write\equa{}}\ewrite%
\else%
\ifnum\nosubsection=0%
\xdef#1{\noexpand\hyperref{}{section}{\prefix\the\nosection}{\prefix\the\nosection}}%
\edef\ewrite{\write\eqdf{\string\def\string#1{\prefix%
\the\nosection}}\write\eqdf{}}\ewrite%
\edef\ewrite{\write\equa{{\string#1},\prefix\the\nosection}%
\write\equa{}}\ewrite%
\writedef{#1\leftbracket#1}
\else%
\xdef#1{\noexpand\hyperref{}{subsection}{\prefix\the\nosection.%
\the\nosubsection}{\prefix\the\nosection.\the\nosubsection}}%
\writedef{#1\leftbracket#1}
\edef\ewrite{\write\eqdf{\string\def\string#1{\prefix%
\the\nosection.\the\nosubsection}}\write\eqdf{}}\ewrite%
\edef\ewrite{\write\equa{{\string#1},\prefix\the\nosection%
.\the\nosubsection}\write\equa{}}\ewrite\fi\fi}%
\def\listcontent{\immediate\closeout\tfile\immediate\openin%
\ch@ckfile=\jobname.tab
\ifeof\ch@ckfile\message{no file \jobname.tab, no table of contents this
pass}%
\else\closein\ch@ckfile\centerline{\bf\iffrancmode Table des
mati\`eres \else Contents\fi}\nobreak\medskip%
{\baselineskip=12pt\parskip=0pt\catcode`\@=11\input\jobname.tab
\catcode`\@=12\bigbreak\bigskip}\fi}
\newwrite\tfile \def\writetoca#1{}
\def\writetoc{\immediate\openout\tfile=\jobname.tab
   \def\writetoca##1{{\edef\next{\write\tfile{\noindent ##1
   \string\leaderfill {\string\hyperref{}{page}{\noexpand\number\pageno}%
                       {\noexpand\number\pageno}} \par}}\next}}}

%
\def\nolabels{\def\wrlabeL##1{}\def\eqlabeL##1{}\def\reflabeL##1{}}
\def\writelabels{\def\wrlabeL##1{\leavevmode\vadjust{\rlap{\smash%
{\line{{\escapechar=` \hfill\rlap{\sevenrm\hskip.03in\string##1}}}}}}}%
\def\eqlabeL##1{{\escapechar-1\rlap{\sevenrm\hskip.05in\string##1}}}%
\def\reflabeL##1{\noexpand\llap{\noexpand\sevenrm\string\string\string##1}}}
\nolabels

\global\newcount\refno \global\refno=1
\newwrite\rfile
\def\ref{[\hyperref{}{reference}{\the\refno}{\the\refno}]\nref}
\def\nref#1{\DefWarn#1%
\xdef#1{[\noexpand\hyperref{}{reference}{\the\refno}{\the\refno}]}%
\writedef{#1\leftbracket#1}%
\ifnum\refno=1\immediate\openout\rfile=\jobname.ref\fi
\chardef\wfile=\rfile\immediate\write\rfile{\noexpand\item{[\noexpand\hyperdef%
\noexpand\hypernoname{reference}{\the\refno}{\the\refno}]\ }%
\reflabeL{#1\hskip.31in}\pctsign}\global\advance\refno by1\findarg}
\def\findarg#1#{\begingroup\obeylines\newlinechar=`\^^M\pass@rg}
{\obeylines\gdef\pass@rg#1{\writ@line\relax #1^^M\hbox{}^^M}%
\gdef\writ@line#1^^M{\expandafter\toks0\expandafter{\striprel@x #1}%
\edef\next{\the\toks0}\ifx\next\em@rk\let\next=\endgroup\else\ifx\next\empty%
\else\immediate\write\wfile{\the\toks0}\fi\let\next=\writ@line\fi\next\relax}}
\def\striprel@x#1{} \def\em@rk{\hbox{}}
\def\lref{\begingroup\obeylines\lr@f}
\def\lr@f#1#2{\DefWarn#1\gdef#1{\let#1=\UNd@FiNeD\ref#1{#2}}\endgroup\unskip}
\def\semi{;\hfil\break}
\def\addref#1{\immediate\write\rfile{\noexpand\item{}#1}} 
\def\listrefs{{}\vfill\supereject\immediate\closeout\rfile\writestoppt
\baselineskip=14pt\centerline{{\bf\iffrancmode R\'eferences\else References%
\fi}}
\bigskip{\parindent=20pt%
\frenchspacing\escapechar=` \input \jobname.ref\vfill\eject}\nonfrenchspacing}
\def\startrefs#1{\immediate\openout\rfile=\jobname.ref\refno=#1}
\def\xref{\expandafter\xr@f}\def\xr@f[#1]{#1}
\def\refs#1{\count255=1[\r@fs #1{\hbox{}}]}
\def\r@fs#1{\ifx\UNd@FiNeD#1\message{reflabel \string#1 is undefined.}%
\nref#1{need to supply reference \string#1.}\fi%
\vphantom{\hphantom{#1}}{\let\hyperref=\relax\xdef\next{#1}}%
\ifx\next\em@rk\def\next{}%
\else\ifx\next#1\ifodd\count255\relax\xref#1\count255=0\fi%
\else#1\count255=1\fi\let\next=\r@fs\fi\next}
%
\newwrite\lfile
{\escapechar-1\xdef\pctsign{\string\%}\xdef\leftbracket{\string\{}
\xdef\rightbracket{\string\}}\xdef\numbersign{\string\#}}
\def\writedefs{\immediate\openout\lfile=\jobname.def \def\writedef##1{%
{\let\hyperref=\relax\let\hyperdef=\relax\let\hypernoname=\relax
 \immediate\write\lfile{\string\def\string##1\rightbracket}}}}%
\def\writestop{\def\writestoppt{\immediate\write\lfile{\string\pageno%
\the\pageno\string\startrefs\leftbracket\the\refno\rightbracket%
\string\def\string\secsym\leftbracket\secsym\rightbracket%
\string\secno\the\secno\string\meqno\the\meqno}\immediate\closeout\lfile}}
\def\writestoppt{}\def\writedef#1{}
\writedefs
\def\biblio\par{\vskip0pt plus.1\vsize\penalty-100\vskip0pt plus-.1
\vsize\bigskip\vskip\parskip
\message{Bibliographie}
{\leftline{\bf \hyperdef\hypernoname{biblio}{bib}{Bibliographical Notes}}}
\nobreak\medskip\noindent\frenchspacing
\writetoca{\string\hyperref{}{biblio}{bib}{Bibliographical Notes}}}%

\def\biblionote{\iffrancmode Notes Bibliographiques\else Bibliographical Notes
\fi}
\def\beginbib\par{\vskip0pt plus.1\vsize\penalty-100\vskip0pt plus-.1
\vsize\bigskip\vskip\parskip
\message{Bibliographie}
{\leftline{\bf \hyperdef\hypernoname{biblio}{\the\nosection}%
{\biblionote}}}
\nobreak\medskip\noindent\frenchspacing
\writetoca{\string\hyperref{}{biblio}{\the\nosection}%
{\biblionote}}}%

\def\Exercises{\iffrancmode Exercices\else Exercises
\fi}
\def\exerc\par{\vskip0pt plus.1\vsize\penalty-100\vskip0pt plus-.1
\vsize\bigskip\vskip\parskip
\message{Exercises}
{\leftline{\bf \hyperdef\hypernoname{exercise}{\the\nosection}{\Exercises}}}
\nobreak\medskip\noindent\frenchspacing
\writetoca{\string\hyperref{}{exercise}{\the\nosection}{\Exercises}}
}
\def\eqnn{\global\advance\neqno by 1 \ifinner\relax\else%
\eqno\fi(\prefix\the\nosection.\the\neqno)}
%
\def\eqnd#1{\global\advance\neqno by 1 
{\xdef#1{($\noexpand\hyperref{}{equation}{\prefix\the\nosection.\the\neqno}%
{\prefix\the\nosection.\the\neqno}$)}}
\ifinner\relax\else\eqno\fi(\hyperdef\hypernoname{equation}{\prefix\the%
\nosection.\the\neqno}{\prefix\the\nosection.\the\neqno})
\writedef{#1\leftbracket#1}
\ifdraftmode{\escapechar-1{\rlap{\hskip.2mm\sevenrm\string#1}}}\fi
\edef\ewrite{\write\eqdf{\string\def\string#1{($\prefix\the\nosection.%
\the\neqno$)}}%
\write\eqdf{}}\ewrite%
\edef\ewrite{\write\equa{{\string#1},(\prefix\the\nosection.\the\neqno)
{\noexpand\number\pageno}}\write\equa{}}\ewrite}
%
\def\checkm@de#1#2{\ifmmode{\def\f@rst##1{##1}\hyperdef\hypernoname{equation}%
{#1}{#2}}\else\hyperref{}{equation}{#1}{#2}\fi}
\def\f@rst#1{\c@t#1a\em@ark}\def\c@t#1#2\em@ark{#1}
\def\eqna#1{\global\advance\neqno by1\ifdraftmode{\hfill%
\escapechar-1{\rlap{\sevenrm\string#1}}}\fi%
\xdef #1##1{(\noexpand\relax\noexpand%
\checkm@de{\prefix\the\nosection.\the\neqno\noexpand\f@rst{##1}1}%
{\hbox{$\prefix\the\nosection.\the\neqno##1$}})}
\writedef{#1\numbersign1\leftbracket#1{\numbersign1}}%
} 
%

%
\def\em@rk{\hbox{}} 
\def\xeqn{\expandafter\xe@n}\def\xe@n(#1){#1}
\def\xeqna#1{\expandafter\xe@na#1}\def\xe@na\hbox#1{\xe@nap #1}
\def\xe@nap$(#1)${\hbox{$#1$}}
\def\eqns#1{(\e@ns #1{\hbox{}})}
\def\e@ns#1{\ifx\UNd@FiNeD#1\message{eqnlabel \string#1 is undefined.}%
\xdef#1{(?.?)}\fi{\let\hyperref=\relax\xdef\next{#1}}%
\ifx\next\em@rk\def\next{}%
\else\ifx\next#1\xeqn#1\else\def\n@xt{#1}\ifx\n@xt\next#1\else\xeqna#1\fi
\fi\let\next=\e@ns\fi\next}
\def\figure#1#2{\global\advance\nofigure by 1 \vglue#1%
{\elevenpoint
\setbox1=\hbox{#2}
\ifdim\wd1=0pt\centerline{Fig.\ \the\nofigure\hskip0.5mm}%
\else\def\caption{Fig.\ \the\nofigure\quad#2\hskip0mm}
\setbox0=\hbox{\caption}
\ifdim\wd0>\hsize\noindent Fig.\ \the\nofigure\quad#2\else
                 \centerline{\caption}\fi\fi}\par}
\def\lfigure#1#2{\global\advance\nofigure by
1\vglue#1\leftline{\elevenpoint\hskip10truemm  Fig.\
\the\nofigure\quad #2}} 
\catcode`@=12

\def\draftend{\vfill\supereject%
\immediate\closeout\equa\immediate\closeout\tab
\ifdraftmode
{\bf \titlename},\par ------------ Date \today. -----------\par
\edef\ewrite{\write\eqdf{}}\ewrite%
\catcode`\&=0
\catcode`\\=10
\input \equation
\catcode`\\=0
\catcode`\&=4\fi
\end
}

%% file: lfont
\def\sla#1{\mkern-1.5mu\raise0.4pt\hbox{$\not$}\mkern1.2mu #1\mkern 0.7mu}
\def\Dbar{\mkern-1.5mu\raise0.4pt\hbox{$\not$}\mkern-.1mu {\rm D}\mkern.1mu}
\def\Abar{\mkern1.mu\raise0.4pt\hbox{$\not$}\mkern-1.3mu A\mkern.1mu}
\def\Bbar{\mkern-0.mu\raise0.4pt\hbox{$\not$}\mkern-.3mu B\mkern 0.6mu}
\newskip\tableskipamount \tableskipamount=8pt plus 3pt minus 3pt
\def\tableskip{\vskip\tableskipamount}

\newdimen\chapskip

\font\ssbx=cmssbx10  

\font\caprm=cmr9
\font\capit=cmti9
\font\capbf=cmbx9
\font\capsl=cmsl9
\font\capmi=cmmi9
\font\capex=cmex9
\font\capsy=cmsy9
\chapskip=17.5mm
\def\makeheadline{\vbox to 0pt{\vskip-22.5pt
\line{\vbox to8.5pt{}\the\headline}\vss}\nointerlineskip}
\font\tenbi=cmmib10 
\font\ninebi=cmmib9
\font\sevenbi=cmmib7 
\font\fivebi=cmmib5
\textfont4=\tenbi
\scriptfont4=\sevenbi
\scriptscriptfont4=\fivebi
\font\headrm=cmr10

\font\sixrm=cmr6
\font\fiverm=cmr5
\font\sixmi=cmmi6
\font\fivemi=cmmi5
\font\sixsy=cmsy6
\font\fivesy=cmsy5
\font\sixbf=cmbx6
\font\fivebf=cmbx5
\skewchar\capmi='177 \skewchar\sixmi='177 \skewchar\fivemi='177
\skewchar\capsy='60 \skewchar\sixsy='60 \skewchar\fivesy='60

\def\elevenpoint{
\textfont0=\caprm \scriptfont0=\sixrm \scriptscriptfont0=\fiverm
\def\rm{\fam0\caprm}
\textfont1=\capmi \scriptfont1=\sixmi \scriptscriptfont1=\fivemi
\textfont2=\capsy \scriptfont2=\sixsy \scriptscriptfont2=\fivesy
\textfont3=\capex \scriptfont3=\capex \scriptscriptfont3=\capex
\textfont\itfam=\capit \def\it{\fam\itfam\capit} 
\textfont\slfam=\capsl  \def\sl{\fam\slfam\capsl} 
\textfont\bffam=\capbf \scriptfont\bffam=\sixbf
\scriptscriptfont\bffam=\fivebf
\def\bf{\fam\bffam\capbf} 
\textfont4=\ninebi \scriptfont4=\sevenbi \scriptscriptfont4=\fivebi
\abovedisplayskip=11pt plus 3pt minus 8pt
\belowdisplayskip=\abovedisplayskip
\smallskipamount=2.7pt plus 1pt minus 1pt
\medskipamount=5.4pt plus 2pt minus 2pt
\bigskipamount=10.8pt plus 3.6pt minus 3.6pt
\normalbaselineskip=11pt
\setbox\strutbox=\hbox{\vrule height7.8pt depth3.2pt width0pt}
\normalbaselines \rm}

%
%

\catcode`\@=11

\font\tenmsa=msam10
\font\sevenmsa=msam7
\font\fivemsa=msam5
\font\tenmsb=msbm10
\font\sevenmsb=msbm7
\font\fivemsb=msbm5
\newfam\msafam
\newfam\msbfam
\textfont\msafam=\tenmsa  \scriptfont\msafam=\sevenmsa
  \scriptscriptfont\msafam=\fivemsa
\textfont\msbfam=\tenmsb  \scriptfont\msbfam=\sevenmsb
  \scriptscriptfont\msbfam=\fivemsb

\def\hexnumber@#1{\ifcase#1 0\or1\or2\or3\or4\or5\or6\or7\or8\or9\or
	A\or B\or C\or D\or E\or F\fi }

\font\teneuf=eufm10
\font\seveneuf=eufm7
\font\fiveeuf=eufm5
\newfam\euffam
\textfont\euffam=\teneuf
\scriptfont\euffam=\seveneuf
\scriptscriptfont\euffam=\fiveeuf
\def\frak{\ifmmode\let\next\frak@\else
 \def\next{\Err@{Use \string\frak\space only in math mode}}\fi\next}
\def\goth{\ifmmode\let\next\frak@\else
 \def\next{\Err@{Use \string\goth\space only in math mode}}\fi\next}
\def\frak@#1{{\frak@@{#1}}}
\def\frak@@#1{\fam\euffam#1}

\edef\msa@{\hexnumber@\msafam}
\edef\msb@{\hexnumber@\msbfam}

\mathchardef\boxdot="2\msa@00
\mathchardef\boxplus="2\msa@01
\mathchardef\boxtimes="2\msa@02
\mathchardef\square="0\msa@03
\mathchardef\blacksquare="0\msa@04
\mathchardef\centerdot="2\msa@05
\mathchardef\lozenge="0\msa@06
\mathchardef\blacklozenge="0\msa@07
\mathchardef\circlearrowright="3\msa@08
\mathchardef\circlearrowleft="3\msa@09
\mathchardef\rightleftharpoons="3\msa@0A
\mathchardef\leftrightharpoons="3\msa@0B
\mathchardef\boxminus="2\msa@0C
\mathchardef\Vdash="3\msa@0D
\mathchardef\Vvdash="3\msa@0E
\mathchardef\vDash="3\msa@0F
\mathchardef\twoheadrightarrow="3\msa@10
\mathchardef\twoheadleftarrow="3\msa@11
\mathchardef\leftleftarrows="3\msa@12
\mathchardef\rightrightarrows="3\msa@13
\mathchardef\upuparrows="3\msa@14
\mathchardef\downdownarrows="3\msa@15
\mathchardef\upharpoonright="3\msa@16

\mathchardef\downharpoonright="3\msa@17
\mathchardef\upharpoonleft="3\msa@18
\mathchardef\downharpoonleft="3\msa@19
\mathchardef\rightarrowtail="3\msa@1A
\mathchardef\leftarrowtail="3\msa@1B
\mathchardef\leftrightarrows="3\msa@1C
\mathchardef\rightleftarrows="3\msa@1D
\mathchardef\Lsh="3\msa@1E
\mathchardef\Rsh="3\msa@1F
\mathchardef\rightsquigarrow="3\msa@20
\mathchardef\leftrightsquigarrow="3\msa@21
\mathchardef\looparrowleft="3\msa@22
\mathchardef\looparrowright="3\msa@23
\mathchardef\circeq="3\msa@24
\mathchardef\succsim="3\msa@25
\mathchardef\gtrsim="3\msa@26
\mathchardef\gtrapprox="3\msa@27
\mathchardef\multimap="3\msa@28
\mathchardef\therefore="3\msa@29
\mathchardef\because="3\msa@2A
\mathchardef\doteqdot="3\msa@2B

\mathchardef\triangleq="3\msa@2C
\mathchardef\precsim="3\msa@2D
\mathchardef\lesssim="3\msa@2E
\mathchardef\lessapprox="3\msa@2F
\mathchardef\eqslantless="3\msa@30
\mathchardef\eqslantgtr="3\msa@31
\mathchardef\curlyeqprec="3\msa@32
\mathchardef\curlyeqsucc="3\msa@33
\mathchardef\preccurlyeq="3\msa@34
\mathchardef\leqq="3\msa@35
\mathchardef\leqslant="3\msa@36
\mathchardef\lessgtr="3\msa@37
\mathchardef\backprime="0\msa@38
\mathchardef\risingdotseq="3\msa@3A
\mathchardef\fallingdotseq="3\msa@3B
\mathchardef\succcurlyeq="3\msa@3C
\mathchardef\geqq="3\msa@3D
\mathchardef\geqslant="3\msa@3E
\mathchardef\gtrless="3\msa@3F
\mathchardef\sqsubset="3\msa@40
\mathchardef\sqsupset="3\msa@41
\mathchardef\vartriangleright="3\msa@42
\mathchardef\vartriangleleft="3\msa@43
\mathchardef\trianglerighteq="3\msa@44
\mathchardef\trianglelefteq="3\msa@45
\mathchardef\bigstar="0\msa@46
\mathchardef\between="3\msa@47
\mathchardef\blacktriangledown="0\msa@48
\mathchardef\blacktriangleright="3\msa@49
\mathchardef\blacktriangleleft="3\msa@4A
\mathchardef\vartriangle="0\msa@4D
\mathchardef\blacktriangle="0\msa@4E
\mathchardef\triangledown="0\msa@4F
\mathchardef\eqcirc="3\msa@50
\mathchardef\lesseqgtr="3\msa@51
\mathchardef\gtreqless="3\msa@52
\mathchardef\lesseqqgtr="3\msa@53
\mathchardef\gtreqqless="3\msa@54
\mathchardef\Rrightarrow="3\msa@56
\mathchardef\Lleftarrow="3\msa@57
\mathchardef\veebar="2\msa@59
\mathchardef\barwedge="2\msa@5A
\mathchardef\doublebarwedge="2\msa@5B
\mathchardef\angle="0\msa@5C
\mathchardef\measuredangle="0\msa@5D
\mathchardef\sphericalangle="0\msa@5E
\mathchardef\varpropto="3\msa@5F
\mathchardef\smallsmile="3\msa@60
\mathchardef\smallfrown="3\msa@61
\mathchardef\Subset="3\msa@62
\mathchardef\Supset="3\msa@63
\mathchardef\Cup="2\msa@64

\mathchardef\Cap="2\msa@65

\mathchardef\curlywedge="2\msa@66
\mathchardef\curlyvee="2\msa@67
\mathchardef\leftthreetimes="2\msa@68
\mathchardef\rightthreetimes="2\msa@69
\mathchardef\subseteqq="3\msa@6A
\mathchardef\supseteqq="3\msa@6B
\mathchardef\bumpeq="3\msa@6C
\mathchardef\Bumpeq="3\msa@6D
\mathchardef\lll="3\msa@6E

\mathchardef\ggg="3\msa@6F

\mathchardef\circledS="0\msa@73
\mathchardef\pitchfork="3\msa@74
\mathchardef\dotplus="2\msa@75
\mathchardef\backsim="3\msa@76
\mathchardef\backsimeq="3\msa@77
\mathchardef\complement="0\msa@7B
\mathchardef\intercal="2\msa@7C
\mathchardef\circledcirc="2\msa@7D
\mathchardef\circledast="2\msa@7E
\mathchardef\circleddash="2\msa@7F
\def\ulcorner{\delimiter"4\msa@70\msa@70 }
\def\urcorner{\delimiter"5\msa@71\msa@71 }
\def\llcorner{\delimiter"4\msa@78\msa@78 }
\def\lrcorner{\delimiter"5\msa@79\msa@79 }
\def\yen{\mathhexbox\msa@55 }
\def\checkmark{\mathhexbox\msa@58 }
\def\circledR{\mathhexbox\msa@72 }
\def\maltese{\mathhexbox\msa@7A }
\mathchardef\lvertneqq="3\msb@00
\mathchardef\gvertneqq="3\msb@01
\mathchardef\nleq="3\msb@02
\mathchardef\ngeq="3\msb@03
\mathchardef\nless="3\msb@04
\mathchardef\ngtr="3\msb@05
\mathchardef\nprec="3\msb@06
\mathchardef\nsucc="3\msb@07
\mathchardef\lneqq="3\msb@08
\mathchardef\gneqq="3\msb@09
\mathchardef\nleqslant="3\msb@0A
\mathchardef\ngeqslant="3\msb@0B
\mathchardef\lneq="3\msb@0C
\mathchardef\gneq="3\msb@0D
\mathchardef\npreceq="3\msb@0E
\mathchardef\nsucceq="3\msb@0F
\mathchardef\precnsim="3\msb@10
\mathchardef\succnsim="3\msb@11
\mathchardef\lnsim="3\msb@12
\mathchardef\gnsim="3\msb@13
\mathchardef\nleqq="3\msb@14
\mathchardef\ngeqq="3\msb@15
\mathchardef\precneqq="3\msb@16
\mathchardef\succneqq="3\msb@17
\mathchardef\precnapprox="3\msb@18
\mathchardef\succnapprox="3\msb@19
\mathchardef\lnapprox="3\msb@1A
\mathchardef\gnapprox="3\msb@1B
\mathchardef\nsim="3\msb@1C
\mathchardef\ncong="3\msb@1D

\mathchardef\varsubsetneq="3\msb@20
\mathchardef\varsupsetneq="3\msb@21
\mathchardef\nsubseteqq="3\msb@22
\mathchardef\nsupseteqq="3\msb@23
\mathchardef\subsetneqq="3\msb@24
\mathchardef\supsetneqq="3\msb@25
\mathchardef\varsubsetneqq="3\msb@26
\mathchardef\varsupsetneqq="3\msb@27
\mathchardef\subsetneq="3\msb@28
\mathchardef\supsetneq="3\msb@29
\mathchardef\nsubseteq="3\msb@2A
\mathchardef\nsupseteq="3\msb@2B
\mathchardef\nparallel="3\msb@2C
\mathchardef\nmid="3\msb@2D
\mathchardef\nshortmid="3\msb@2E
\mathchardef\nshortparallel="3\msb@2F
\mathchardef\nvdash="3\msb@30
\mathchardef\nVdash="3\msb@31
\mathchardef\nvDash="3\msb@32
\mathchardef\nVDash="3\msb@33
\mathchardef\ntrianglerighteq="3\msb@34
\mathchardef\ntrianglelefteq="3\msb@35
\mathchardef\ntriangleleft="3\msb@36
\mathchardef\ntriangleright="3\msb@37
\mathchardef\nleftarrow="3\msb@38
\mathchardef\nrightarrow="3\msb@39
\mathchardef\nLeftarrow="3\msb@3A
\mathchardef\nRightarrow="3\msb@3B
\mathchardef\nLeftrightarrow="3\msb@3C
\mathchardef\nleftrightarrow="3\msb@3D
\mathchardef\divideontimes="2\msb@3E
\mathchardef\varnothing="0\msb@3F
\mathchardef\nexists="0\msb@40
\mathchardef\mho="0\msb@66
\mathchardef\eth="0\msb@67
\mathchardef\eqsim="3\msb@68
\mathchardef\beth="0\msb@69
\mathchardef\gimel="0\msb@6A
\mathchardef\daleth="0\msb@6B
\mathchardef\lessdot="3\msb@6C
\mathchardef\gtrdot="3\msb@6D
\mathchardef\ltimes="2\msb@6E
\mathchardef\rtimes="2\msb@6F
\mathchardef\shortmid="3\msb@70
\mathchardef\shortparallel="3\msb@71
\mathchardef\smallsetminus="2\msb@72
\mathchardef\thicksim="3\msb@73
\mathchardef\thickapprox="3\msb@74
\mathchardef\approxeq="3\msb@75
\mathchardef\succapprox="3\msb@76
\mathchardef\precapprox="3\msb@77
\mathchardef\curvearrowleft="3\msb@78
\mathchardef\curvearrowright="3\msb@79
\mathchardef\digamma="0\msb@7A
\mathchardef\varkappa="0\msb@7B
\mathchardef\hslash="0\msb@7D
\mathchardef\hbar="0\msb@7E
\mathchardef\backepsilon="3\msb@7F
\def\Bbb{\ifmmode\let\next\Bbb@\else
 \def\next{\errmessage{Use \string\Bbb\space only in math mode}}\fi\next}
\def\Bbb@#1{{\Bbb@@{#1}}}
\def\Bbb@@#1{\fam\msbfam#1}
\font\sacfont=eufm10 scaled 1440
\catcode`\@=12